\documentclass[%
 reprint,
 superscriptaddress,
 amsmath,amssymb,
 aps,
]{revtex4-2}

\usepackage{graphicx}% Include figure files
\usepackage[OliveGreen,dvipsnames]{xcolor}
\usepackage{dcolumn}% Align table columns on decimal point
\usepackage{bm}% bold math
\definecolor{blue(ryb)}{rgb}{0.01, 0.28, 1.0}

\newcommand{\MB}[1]{{\color{black} #1}}

\begin{document}

\preprint{APS/123-QED}

\title{Transition route to elastic and elasto-inertial turbulence in polymer channel flows}

\author{M. Beneitez}
\email{mb2467@cam.ac.uk}
\affiliation{
 DAMTP, Centre for Mathematical Sciences, Wilberforce Road, Cambridge CB3 0WA, UK
} 

\author{J. Page}%
\email{jacob.page@ed.ac.uk}
\affiliation{
 School of Mathematics, University of Edinburgh, EH9 3FD, UK
}

\author{Y. Dubief}
\email{ydubief@uvm.edu}
\affiliation{
 Department of Mechanical Engineering, University of Vermont, Burlington, VT, USA
}

\author{R. R. Kerswell}
\email{rrk26@cam.ac.uk}
\affiliation{
 DAMTP, Centre for Mathematical Sciences, Wilberforce Road, Cambridge CB3 0WA, UK
}

\date{\today}% It is always \today, today,

\begin{abstract}

Viscoelastic shear flows support additional chaotic states beyond simple Newtonian turbulence. In vanishing Reynolds number flows, the nonlinearity \MB{in the polymer evolution equation alone can sustain} inertialess `elastic’ turbulence (ET) while `elasto-inertial’ turbulence (EIT) appears to rely on an interplay between elasticity and finite-$Re$ effects. Despite their distinct phenomenology and industrial significance, transition routes and possible connections between these states are unknown. 
\MB{We identify here a common Ruelle-Takens transition scenario for both of these chaotic regimes in two-dimensional direct numerical simulations of FENE-P fluids in a straight channel. The primary bifurcation is caused by a recently-discovered `polymer diffusive instability' associated with small but non-vanishing polymer stress diffusion which generates a finite-amplitude, small-scale travelling wave localised at the wall. 
%
% Beneitez '23, Couchman '24 and Lewy '24 all use Polymer DiffusIVE Instability
%
This is found to be unstable to a large-scale secondary instability which grows to modify the whole flow before itself breaking down in a  third bifurcation to either ET or EIT.}
The secondary large-scale instability waves resemble `centre’ and `wall’ modes respectively — instabilities which have been conjectured to play a role in viscoelastic chaotic dynamics but were previously \MB{only} thought to exist far from relevant areas of the parameter space. 

\end{abstract}

\maketitle

Viscoelastic polymer solutions are found in a wide variety of industrial and biological flows and often exhibit striking counter-intuitive behaviours (e.g. rod climbing, elastic recoil \cite{datta2022perspectives,sanchez22}). 
More importantly, there are new chaotic flow states which can be sustained in viscoelastic shear flows beyond a perturbed version of Newtonian turbulence. 
In flows with vanishing inertia, polymer solutions can sustain ``elastic turbulence'' (ET) \cite{groisman2000elastic} which is maintained by nonlinearities in the polymer stress \MB{equation}, while ``elasto-inertial turbulence'' (EIT) apparently requires both elastic and inertial effects \cite{samanta2013elasto,dubief2023elasto}. 
Despite much recent progress including the \MB{recent discovery of a \MB{centre-mode} linear instability} \cite{garg2018viscoelastic,khalid2021centre,khalid2021continuous}, exact nonlinear solutions \cite{page2020exact,buza2022finite,morozov2022coherent} and direct numerical simulations of chaotic dynamics in planar geometries \cite{beneitez2023polymer,lellep2024purely}, it is still unknown as to how a smooth laminar state might transition to viscoelastic turbulence at realistic parameter settings, and whether ET and EIT are dynamically connected to one another. 

ET was originally discovered in curved geometries where streamline curvature provides a mechanism for the growth of ``hoop stress'' linear instabilities \cite{larson1990purely,pakdel1996elastic}, although experimental evidence \cite{pan2013nonlinear}, supported by recent nonlinear simulations \cite{lellep2024purely} show that ET can be sustained in planar flows in which these instabilities are absent from the basic state. 
Similarly, EIT is also observed in planar channel flows \cite{samanta2013elasto,sid2018two} in the \MB{apparent} absence of linear instability, with the viscoelastic version of the classical Tollmien-Schlicting (TS) wave \cite{zhang2013linear} and the centre mode instability \cite{garg2018viscoelastic,khalid2021centre,khalid2021continuous} existing in regions of the parameter space far from the chaotic dynamics. 
While recent numerical results \cite{lellep2023linear,lellep2024purely} connect ET to a 3D instability of a 2D finite-amplitude `arrowhead’ travelling wave, linking EIT to an exact coherent state has been challenging \cite{shekar2019critical,shekar2021tollmien,beneitez2024multistability}, and a route to either form of chaos from the laminar solution remains an open question. \MB{One new possibility is the very recently discovered `polymer diffusive instability’ (PDI) \cite{beneitez2023polymer,couchman2024inertial,lewy24}, though this is small scale and localised at the walls while observations of both ET and EIT show large scale structures in the interior}. 

% P3 purpose 
The purpose of this Letter is to show that PDI \MB{does indeed} provide a mechanism for transition from smooth laminar flow to both ET and EIT in two-dimensional channel flows of FENE-P fluids. 
The small-scale linear instability of the laminar flow leads to the formation of a low-amplitude, wall-localised travelling wave, which we show is itself linearly unstable to large-scale secondary instabilities with wavelength on the order of the channel height which fill the flow's core. 
In ET this \MB{second} linear instability resembles the `centre mode’ whereas in EIT it resembles a perturbed version of a viscoelastic TS wave (or `wall mode'). 
Our results thereby establish a common pathway to both chaotic states in viscoelastic flows, and suggest
that other small-scale perturbations (e.g. wall roughness \cite{Page2022}) may also provide the seed for these secondary instabilities. 

%
% fig 1
%
\begin{figure}
    \centering
    \begin{tabular}{cc}
       \includegraphics[width=0.665\columnwidth]{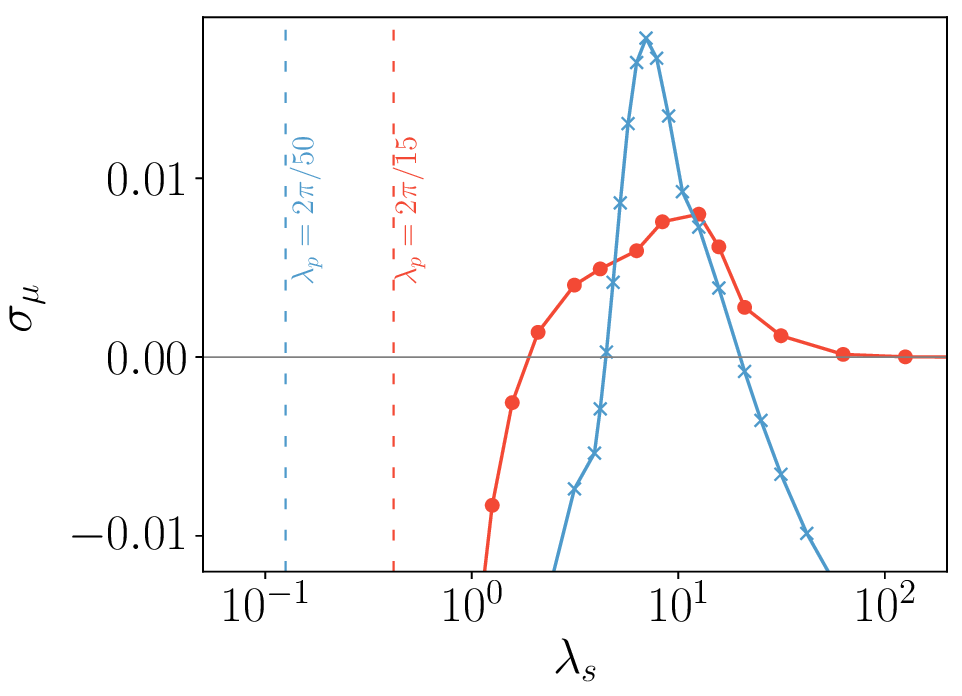}  & \includegraphics[width=0.323\columnwidth]{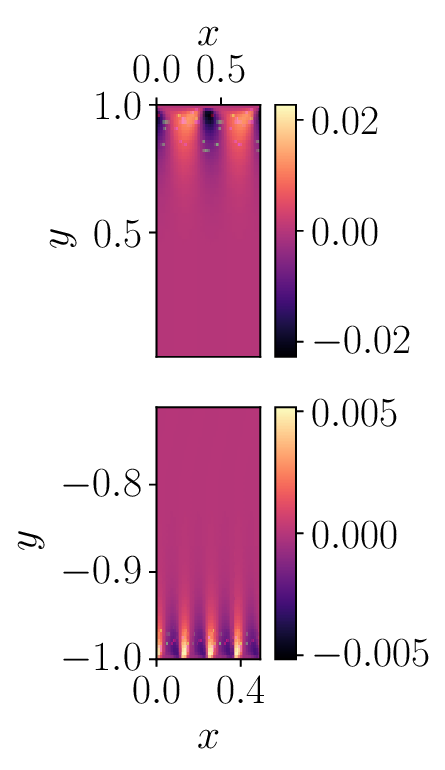}
    \end{tabular}
    \caption{
    Left: Growth rate, $\sigma_{\mu}$, for the secondary instability with respect to the secondary instability lengthscale $\lambda_s$ for PDI at $\mathcal{S}_1$ (blue line, markers indicate Floquet calculations) and $\mathcal{S}_2$ (red line, markers indicate Floquet calculations). Vertical lines denote the lengthscale of the primary PDI instability for $\mathcal{S}_1$ (blue) and $\mathcal{S}_2$ (red).
    Top right: wall-normal velocity $v$ contours  for two copies of the base flow of the secondary instability corresponding to the parameter set $\mathcal{S}_1$. Bottom right: wall-normal velocity $v$ contours for three copies base flow corresponds to the parameter set $\mathcal{S}_2$.}
    \label{fig:lscale_floquet_EIT}
\end{figure}

We consider two-dimensional, pressure-driven channel flow of a dilute polymer solution between two walls separated by a distance $2h$. The polymer contributes an extra stress in the momentum equation, which together with conservation of mass read
\begin{subequations}
\begin{align}
    \hspace{-0.1cm}\textit{Re} (\partial_t \mathbf{u} + \mathbf{u}\cdot \nabla \mathbf{u}) + \nabla p &= \beta\Delta \mathbf{u}+(1-\beta)\nabla \cdot \mathbf{T}(\mathbf{C}),\label{eq:Ueq}\\
    \nabla \cdot \mathbf{u} &= 0. \label{eq:divFree}
\end{align}
Assuming a FENE-P fluid, where the polymeric stress is related to the conformation tensor $\mathbf C$ - an ensemble average of the product of the end-to-end vector of each polymer molecule - via
\begin{equation}    
\mathbf{T}(\mathbf{C}) := \frac{1}{\textit{Wi}}\left(
\frac{\mathbf{C}}{1-(\text{tr}(\mathbf{C})-3)/L_{\text{max}}^2}-\mathbf{I}
\right).
\end{equation}
The conformation tensor evolves according to
\begin{equation}
    \partial_t \mathbf{C}+ (\mathbf{u}\cdot \nabla ) \mathbf{C} + \mathbf{T}(\mathbf{C}) = \mathbf{C}\cdot \nabla \mathbf{u} + (\nabla \mathbf{u})^T\cdot \mathbf{C}+ \varepsilon \Delta \mathbf{C}. \label{eq:Ceq}
\end{equation}
\label{eq:eqs}
\end{subequations}
In these equations $\mathbf{u}=(u,v)$ is the velocity with $u$ and $v$ the streamwise and wall-normal velocity respectively, $p$ is the pressure, $\beta:=\nu_s/\nu$ is a ratio of kinematic viscosities, where $\nu_s$ and $\nu_p=\nu-\nu_s$ are the solvent and polymer contributions respectively, and $L_{\text{max}}$ is the maximum extensibility of the polymer chains. The equations have been made non-dimensional with the half-distance between the plates $h$ and the bulk velocity $U_b$ which is kept constant by a time-varying pressure gradient. The remaining non-dimensional parameters are the Reynolds, $\textit{Re}:=U_bh/\nu$, and Weissenberg, $\textit{Wi}:=\tau  U_b/h$, numbers, where $\tau$ is the polymer relaxation time, along with the parameter $\varepsilon := D / U_b h$ which is the dimensionless polymer stress diffusivity.
%
% bcs and parameters
%
For non-zero \MB{polymer stress diffusivity (or centre-of-mass diffusivity)} $\varepsilon \neq 0$ equation (\ref{eq:Ceq}) requires boundary conditions. 
Here we follow convention \cite{sureshkumar1997direct} and apply equation (\ref{eq:Ceq}) with $\varepsilon=0$ at the walls.
The presence of a non-zero $\varepsilon$ introduces a linear instability \MB{\cite{beneitez2023polymer}} which is relatively insensitive to the choice of boundary conditions on $\mathbf C$ \cite{beneitez2023polymer} and retains a finite growth rate as $\varepsilon \rightarrow 0$ \cite{couchman2024inertial}.
This is the polymer diffusive instability (PDI) discussed above, which is a wall mode localised in boundary layers of thickness $\delta \sim \varepsilon^{1/2}$ and has  streamwise wavenumber $k\sim \varepsilon^{-1/2}$. 
Since realistic values of $\varepsilon$ are very small (typically ranging from $\varepsilon\sim 10^{-3}$ for short polymer molecules to $\varepsilon\sim 10^{-6}$ for longer chains \cite{hiemenz2007polymer}) this instability is associated with very short streamwise lengthscales. 
The instability is present both in inertialess and inertia-dominated channel flow \cite{couchman2024inertial} \MB{as well as highly concentrated polymer solutions  \cite{lewy24}.}

In this paper we consider two parameter settings at which PDI is operational: one relevant to elasto-inertial flow $\mathcal S_1:= (Re^{(1)}=1000, Wi^{(1)}=30, \beta^{(1)}=0.7, L_{\text{max}}^{(1)}=200, \varepsilon^{(1)}=4 \times 10^{-5}, k_p^{(1)}=50)$, and one relevant to inertialess elastic turbulence,  $\mathcal S_2:= (Re^{(2)}=0, Wi^{(2)}=7, \beta^{(2)}=0.7, L_{\text{max}}^{(2)}=200, \varepsilon^{(2)}=5 \times 10^{-4}, k_p^{(2)}=15)$. In these expressions $k_p$ is the streamwise wavenumber at which (primary) PDI is excited. We report full details of the primary instability and its time evolution in the Supplemental Information (SI).  In what follows, the focus is on the secondary dynamics which result from this primary instability saturating at a small but finite amplitude at the walls.

%
% fig 2
%
\begin{figure*}
    \centering 
    \begin{minipage}{0.99\columnwidth}
    \includegraphics[width=1.02\columnwidth]{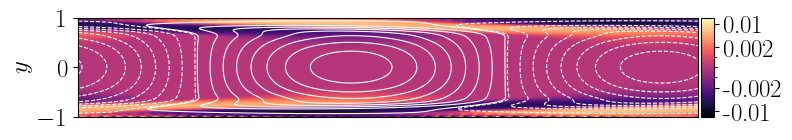} \\
    \includegraphics[width=1.02\columnwidth]{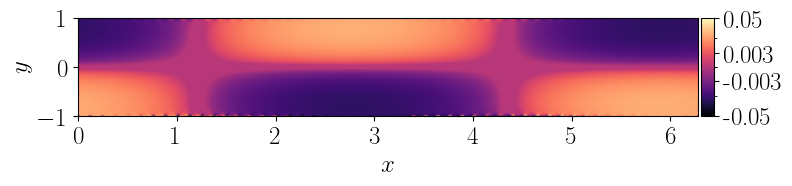}
    \end{minipage}
    \begin{minipage}{0.99\columnwidth}
    \includegraphics[width=1.02\columnwidth]{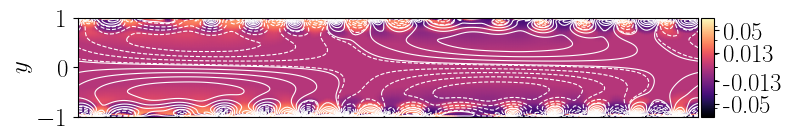}\\
    \includegraphics[width=1.02\columnwidth]{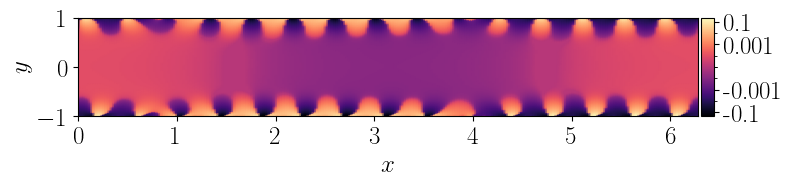}
    \end{minipage}
    %\put(-10, 50){$\times 10^{-3}$} \put(-255, 50){$\times 10^{-2}$}
    \caption{Left: Unstable Floquet mode for $k_s=1$ for parameter set $\mathcal S_1$ (top shows $10^{2}\text{tr}({\bf C})/L^2_{\text{max}}$ and bottom pressure $p$). 
    Right: Unstable Floquet mode for $k_s=1$ for parameter set $\mathcal S_2$ 
    (top shows $10^{3}\text{tr}({\bf C})/L^2_{\text{max}}$ and bottom pressure $p$). Lines corresponds to level sets of the streamfunction (solid positive, dashed negative).  The contour plots use a logarithmic scale to show the range of scales present. Note that axes are not to scale.}
    \label{fig:emode_floquet}
\end{figure*}

%
% Calculations
%

We initialise a calculation for each parameter set on a computational domain of length $\lambda_p^{(i)}:=2\pi / k_p^{(i)}$, with the laminar flow perturbed by $O(10^{-11})$ amplitude noise.
In time, this instability saturates into a wall-localised  travelling wave  with phase speed $c$ (see figure \ref{fig:lscale_floquet_EIT} where the boundary \MB{layer thickness} is  $O(\varepsilon^{1/2})$ ) which is stable in these small computational domains (full details in SI). We now show that this weakly perturbed laminar flow is the launching point for secondary instabilities with wavelengths comparable to the channel height which can themselves break down to either ET or EIT in longer domains. As the PDI-perturbed laminar flow is spatially periodic and steady in a frame moving at $c{\bf \hat{x}}$, the secondary linear stability analysis must consider general  (Floquet) disturbances of the form
\begin{equation}
    \phi(X,y, t) = \left(\sum_{n=-N_x/2}^{N_x/2} \hat{\phi}_{n}(y)e^{i n k_p X} \right)e^{i(k_s X  -\omega_{\mu} t)} + \text{c.c.}
    \label{eq:floquet_ans}
\end{equation}
where $X:=x-ct$, $\phi$ is a perturbation variable in physical space, $\lambda_s:=2\pi/k_s$  is the modulation wavelength of the secondary instability (with $0<k_s<k_p/2$ the wavenumber) and $\omega_{\mu} = \omega_{\mu}^r + i \sigma_{\mu}$ is the complex frequency of the perturbations with $\sigma_{\mu}$ the growth rate. Substituting into the equations linearised about the PDI-perturbed laminar flow in a Galilean frame moving at $c {\bf \hat{x}}$ leads to a generalised eigenvalue problem for $\omega_{\mu}$ with square matrices of side size $7(N_x+1)N_y$, where $N_y$ is the number of Chebyshev polynomials used to discretize the problem in the wall-normal direction and 7 the number of flow variables. \MB{The various computations have different resolution requirements, ranging from $[N_x,N_y]=[12,128]$ for the small-domain computations and the Floquet analysis calculations up to $[N_x,N_y]=[800,800]$ for the large-domain simulations of chaotic states. Each of the following paragraphs includes a detailed description of the required resolution.} The resolution used to resolve the PDI-perturbed laminar state (illustrated in figure \ref{fig:lscale_floquet_EIT}) was $[N_x,N_y]=[12,128]$ modes, with the higher value $[N_x,N_y]=[24,256]$ used to verify the results. The growth rate of the secondary instability for both configurations $\mathcal S_1$ and $\mathcal S_2$ is reported as a function of the modulation wavelength, $\lambda_s$, in figure \ref{fig:lscale_floquet_EIT} where the vertical dashed lines identify the wavelength of the primary PDI, $\lambda_p$.
In both cases there is a range of scales over which secondary instability can occur. Crucially, these scales are $O(1)$ (i.e. on the scale of the channel height) and so not apparently related to the primary PDI scale. The inertialess case $\textit{Re}=0$ (parameter set $\mathcal S_2$) is secondary unstable for $2/3\pi \lesssim \lambda_s \lesssim 20\pi$, with the maximum growth rate at $\lambda_s \sim 4\pi$. In contrast, the range of unstable modulation wavelengths at parameter setting $\mathcal S_1$ is smaller, spanning $2\pi\lesssim \lambda_s \lesssim 4\pi$.

Unstable eigenfunctions for both parameter sets are reported in figure \ref{fig:emode_floquet}. In the elasto-inertial case ($\mathcal S_1$) the instability wave consists of domain-filling rolls with the streamfunction approximately symmetric about the midplane and nearly antisymmetric sheets of $\text{tr}(\mathbf C$) in the near-wall regions (the base flow is not strictly symmetric about the midplane). The appearance is similar to a classical TS wave which has previously been conjectured to play a role in EIT \cite{shekar2021tollmien}. In contrast, the inertialess elastic configuration ($\mathcal S_2$) has an unstable eigenfunction which is approximately antisymmetric in $v$ with respect to the midplane, which is a characteristic shared by the centre-mode instability \cite{garg2018viscoelastic}. 

In both cases the secondary instability waves fill the domain, in contrast to the primary PDI which is small-scale and wall-localised. We now demonstrate a route from the laminar state to EIT and ET, via these secondary instabilities, for $\mathcal S_1$ and $\mathcal S_2$ respectively, using direct numerical simulations.

% Fig 3 time series and snapshots
\begin{figure*}
    \centering
    \begin{minipage}{.7\columnwidth}
        \centering
        
        \includegraphics[width=1.0\columnwidth]{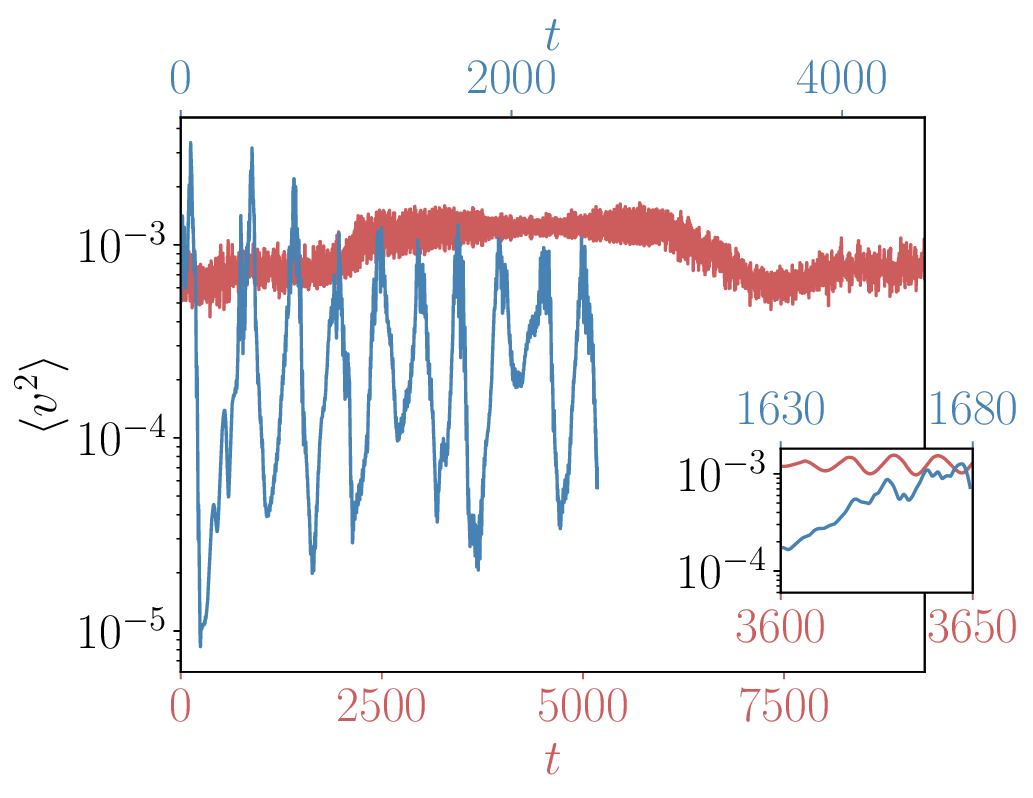}
    \end{minipage}%
    \begin{minipage}{1.25\columnwidth}
        \centering

        \includegraphics[width=1.0\columnwidth]{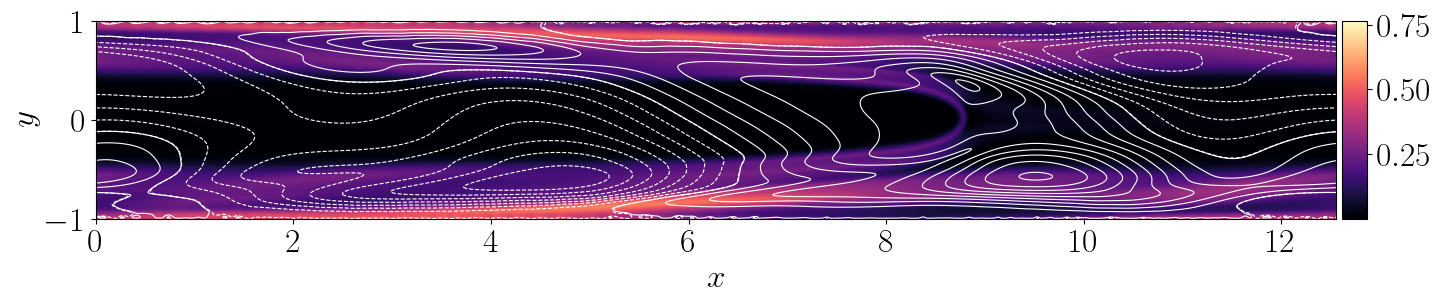}\\
        \includegraphics[width=1.012\columnwidth]{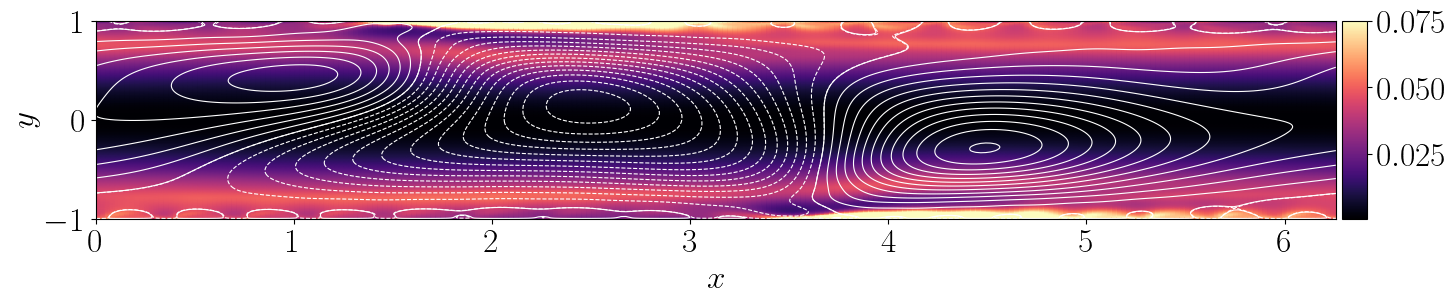}
    \end{minipage}%
    \caption{Left: Time series of the
    \MB{volume-averaged squared normal velocity}
    for EIT (parameter set $\mathcal{S}_1$) in blue and ET (parameter set $\mathcal{S}_2$) in red. Top right: Snapshot of the $\text{tr}({\bf C})/L^2_{\text{max}}$ for the $\textit{Re}=1000$ case. A benign arrowhead structure appears alongside EIT for sufficiently long times. Bottom right: Snapshot of the $\text{tr}({\bf C})/L^2_{\text{max}}$ for the $\textit{Re}=0$ case. Lines corresponds to level sets of the streamfunction (solid positive, white negative). Note axis are not scaled.
    }
    \label{fig:ET_EIT_dns}
\end{figure*}

Starting with the parameter set $\mathcal{S}_1$, the saturated PDI travelling wave on the fundamental domain $2\pi/k_0 = 2\pi/50$ is used to fill \MB{a longer domain of length $L_x=2\pi$ (using resolution $[N_x,N_y] = [256,512]$ and checked with $[N_x,N_y] = [512,512]$). White noise with amplitude $O(10^{-11})$ is then added exciting the unstable secondary instability with wavelength $\lambda_s = 2\pi$ and growth rate in agreement with  Floquet theory. The unstable secondary wave saturates into a quasiperiodic state which resembles a superposition of two wall-localised travelling waves. Two copies of this state were then used as an initial condition in a domain $L_x=4\pi$ to which again white noise with amplitude $O(10^{-11})$
was added. In this longer domain, the quasiperiodic state  is unstable and the flow transitions to EIT in a way reminiscent of Ruelle-Takens  \cite{ruelle1971nature} (this simulation was checked using a higher resolution $[N_x,N_y] = [800,800]$ and also by directly disturbing the saturated PDI state in a $L_x=4\pi$ domain by noise).}
A time series of the volume-averaged kinetic energy of the wall-normal velocity,
$\langle v^2\rangle = \int v^2 \ dV/ (L_x L_y)$, of the chaotic evolution and a snapshot of the flow are reported in figure \ref{fig:ET_EIT_dns}. The contours of $\text{tr}(\mathbf C)$ indicate the presence of stretched polymer `sheets’ in the vicinity of the walls — a feature commonly associated with EIT \cite{dubief2023elasto} — as well as a weak arrowhead structure in the centre of the domain \cite{beneitez2024multistability}. 
%
% fig 4 time series and mean quantities
%
\begin{figure}
    \centering
    \begin{tabular}{cc}
       \includegraphics[width=0.445\columnwidth]{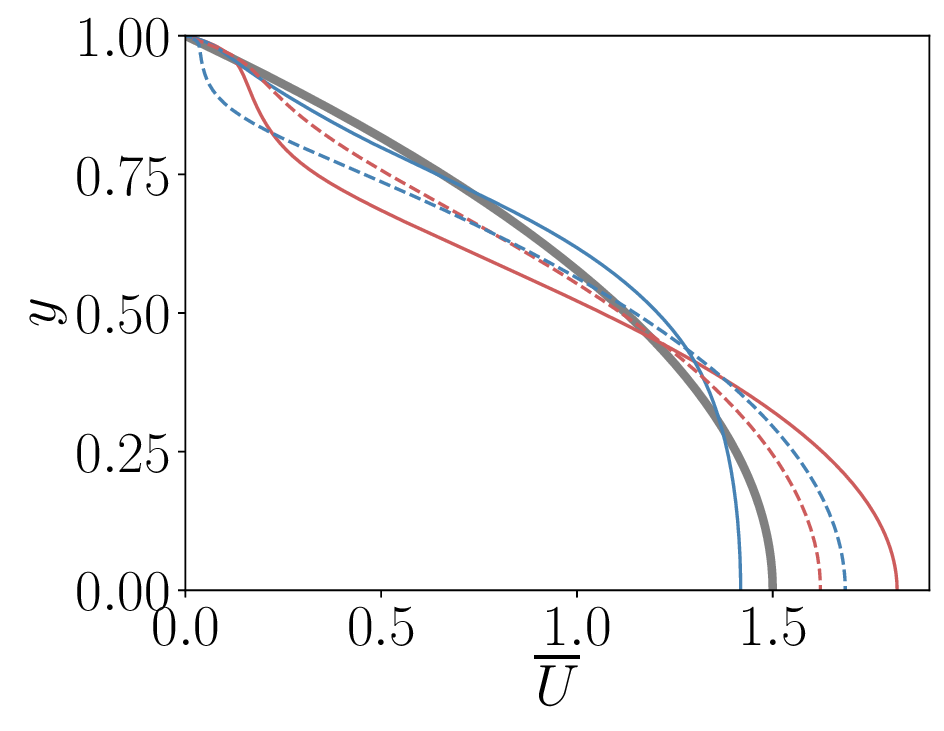}  &  \includegraphics[width=0.56\columnwidth]{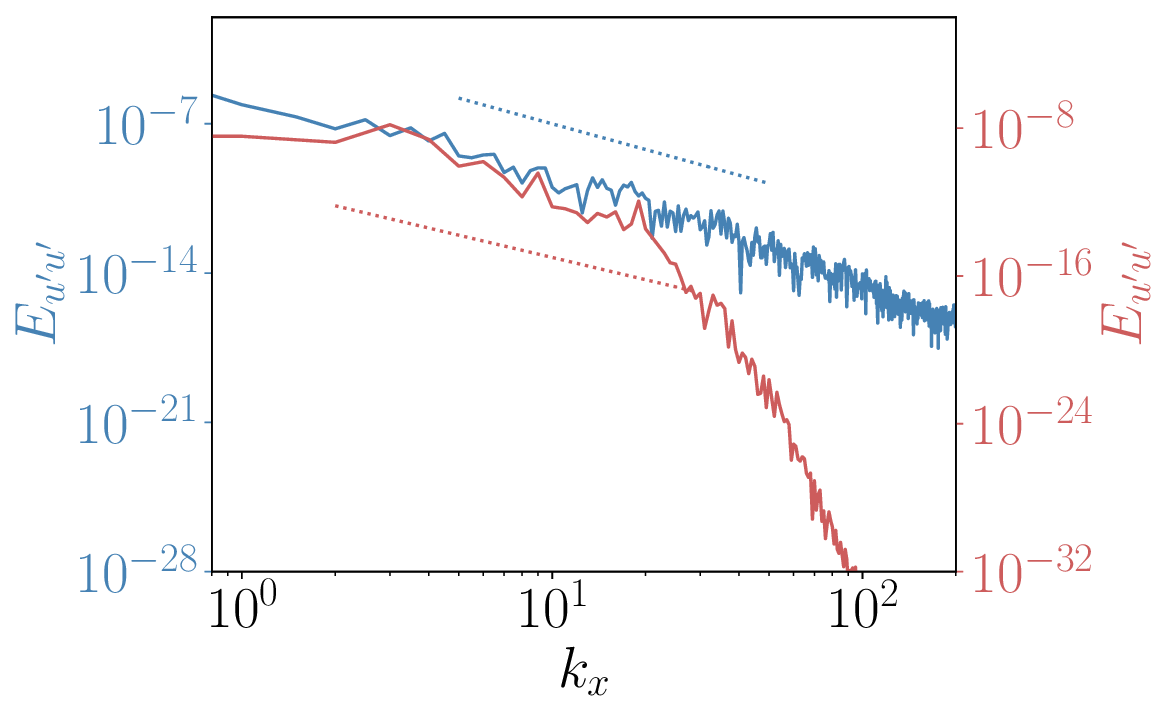} \\
        \includegraphics[width=0.445\columnwidth]{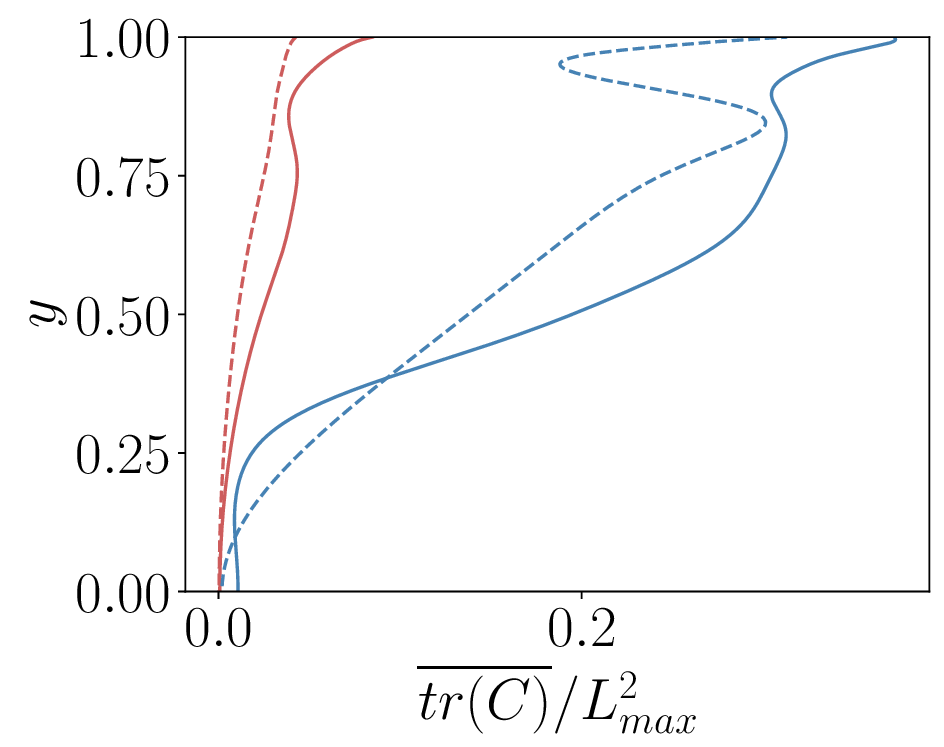} &  \includegraphics[width=0.543\columnwidth]{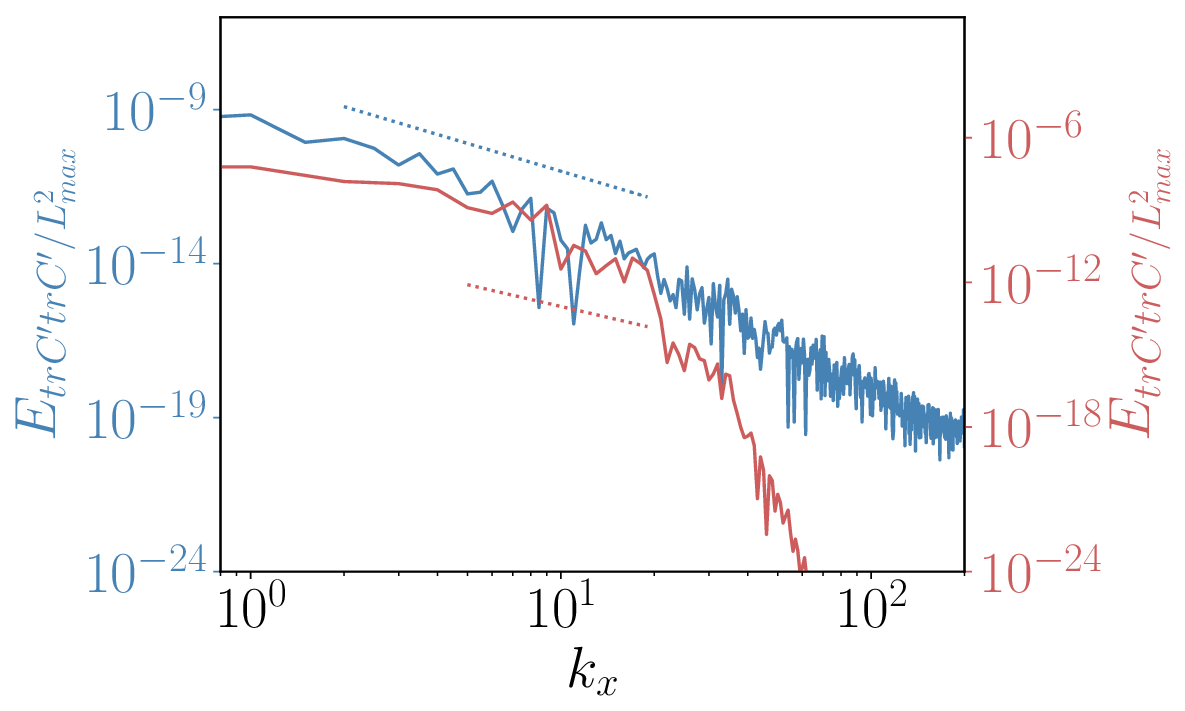}
    \end{tabular}
    \caption{Top left: Mean velocity profile for parameter set $\mathcal{S}_1$ corresponding to: chaotic dynamics (solid blue), nonlinearly saturated PDI (dashed blue), and for parameter set $\mathcal{S}_2$ corresponding to: chaotic dynamics (solid red), nonlinearly saturated PDI (dashed red). The thick grey profile corresponds to the laminar Newtonian flow. Bottom left: idem for the mean trace profile. Top right: Spectrum for the streamwise velocity perturbations at $y=-0.7$ ET (solid red) and EIT case above (solid blue). \MB{Dotted lines indicate slopes of $k_x^{-4}$ in the ET axis (dotted red) and $k_x^{-4}$ in the EIT axis (dotted blue)}. Bottom right: Idem for the \MB{elastic energy} for ET (solid red), and EIT (solid blue). Dotted lines indicate slopes of $k_x^{-3}$ in the ET axis (dotted red) and $k_x^{-3}$ in the EIT axis (dotted blue).}
    \label{fig:ET_EIT_avg}
\end{figure}

A similar procedure is used to trigger a transition to ET in parameter set $\mathcal S_2$. We first take the saturated travelling wave reached from the primary PDI, wavelength $2\pi/k_0 = 2\pi/15$, and tile it over a domain of length $L_x=2\pi$ with a resolution of $[N_x,N_y] = [256,512]$. Perturbing this state with $O(10^{-11})$ white noise excites the unstable Floquet mode with modulation wavelength $\lambda_s=2\pi$, which again grows in accordance with the linear results of figure \ref{fig:lscale_floquet_EIT}, before breaking down to two-dimensional ET in the channel. 
A time series and an example snapshot from the ET evolution are reported in figure \ref{fig:ET_EIT_dns}. 
The flow is dominated by domain-filling vortices in the perturbation velocity, with sheets of polymer extension observed close to the boundaries. 
These flow structures are all found on length scales commensurate with the channel height, although the signature of the primary PDI (which has a longer wavelength than the EIT example) can still be seen in the contour plots. 

Mean-flow statistics for EIT and ET reached in parameter sets $\mathcal S_1$ and $\mathcal S_2$ are reported in figure \ref{fig:ET_EIT_avg}. The time- and streamwise-averaged profiles of the streamwise velocity, $\bar{U}$, show that the mean shear at the wall in both cases is increased relative to the Newtonian laminar flow. At the centreline, EIT exhibits a smaller mean velocity compared to the Newtonian case, while ET has a larger centreline mean velocity. The mean profiles for for the polymer stretch, $\text{tr}(\mathbf C)/L_{\text{max}}^2$, 
indicate that, for these parameter settings, ET features a significantly weakened stretching of the polymer relative to EIT.
The right panels of figure \ref{fig:ET_EIT_avg} report (streamwise) spectra at a particular dynamically relevant \cite{beneitez2024multistability} $y$-location for kinetic energy and perturbations to $\text{tr} (\mathbf C)$. For both ET and EIT the energy spectra \MB{of the streamwise perturbations} show decay \MB{$\propto k_x^{-4}$} for about a decade, with the smaller scales in the inertialess case decaying much faster. A similar behaviour is observed for the perturbations of 
\MB{the elastic energy, defined as 
$E_{E}=-L_{max}^2/(2\textit{Wi}\,\textit{Re}) \ln(1-C_{kk}/L_{max}^2)$ for the inertial case
and as 
$E_{E}=-L_{max}^2/(2\textit{Wi}) \ln(1-C_{kk}/L_{max}^2)$ 
%
%
% L -> L_{max}   
%
in the absence of inertia,} where perturbations for both ET and EIT decay like $\sim k_x^{-3}$ although this is admittedly tentative for ET. The different behaviour for the smaller scales can be attributed to the fact that $\mathcal{S}_1$ has a diffusivity which is roughly an order of magnitude smaller than $\mathcal{S}_2$. \MB{The reported slopes for the spectra in the present work are within the range of slopes observed in the literature \cite{berti2008,berti2010,rota2023elastic}}. 

In this paper we have established a common transition route to both inertialess elastic turbulence (ET) and elasto-inertial turbulence (EIT) in two-dimensional channel flows of FENE-P fluids.
The transition is initiated with a small-scale linear instability which is associated with polymer stress diffusion. 
The resulting wall-localised travelling waves were shown via a Floquet analysis to be unstable to large-scale secondary instabilities resembling either classical TS waves for EIT or a centre-mode instability for ET -- similar instabilities exist for the smooth laminar flow but far from the parameter regime of interest. 
These growing secondary instabilities \MB{generate flows which themselves} become unstable leading to a transition to a chaotic state dominated by large-scale structures in both regimes. 

\MB{Experimentally, it would be hard to observe the first stage of this scenario directly. The primary (PDI) instability is small scale, low intensity and wall-confined where laboratory measurements are difficult. However, the appearance of large scale (secondary) instabilities in parts of parameter space where they are theoretically stable has been observed (e.g. Chouieri \& Hof 2021 who find the centre mode instability in pipe flow at a Reynolds number 10 times smaller than theoretical predictions). While this can be attributed to difficulties in modelling the polymer solution, the present work highlights the importance of wall boundary conditions.
The centre mode instability, at least, is known to be both substantially subcritical (Page et al. 2020) and easily triggered by finite but small disturbances (e.g. fig1b in Buza 2022b) to the extent it could appear as a linear instability in the laboratory (e.g. Chouieri \& Hof 2021, Shnapp \& Steinberg 2022).  It is then not surprising that a small adjustment in the effective boundary conditions, caused specifically here by PDI, can actually make the centre mode linearly unstable.}

\MB{A constant small value $\varepsilon$ of the polymer stress diffusion has been assumed in the analysis, just as one value of the polymer relaxation time $\tau$ is invariably adopted. Prior work \cite{beneitez2023polymer}, however, has found that as long as $\varepsilon$ is small, the PDI growth rate and thus its saturation strength are largely independent of $\varepsilon$. The lengthscale of the instability (boundary layer size and streamwise wavelength) does vary but since the large-scale secondary instability only feels the homogenised (streamwise-averaged) structure of the saturated PDI, this too should be insensitive to the precise value of $\varepsilon$ used.}

Finally, it is worth remarking that the ET obtained here (defined as a chaotic state at vanishing inertia) is two dimensional whereas experimentally-realised ET is three dimensional. Exploring how these states relate to each other is an obvious next challenge.

The authors gratefully acknowledge the support of EPSRC through grant EP/V027247/1.

\bibliography{apssamp}% Produces the bibliography via BibTeX.

\providecommand{\noopsort}[1]{}\providecommand{\singleletter}[1]{#1}%
\begin{thebibliography}{31}%
\makeatletter
\providecommand \@ifxundefined [1]{%
 \@ifx{#1\undefined}
}%
\providecommand \@ifnum [1]{%
 \ifnum #1\expandafter \@firstoftwo
 \else \expandafter \@secondoftwo
 \fi
}%
\providecommand \@ifx [1]{%
 \ifx #1\expandafter \@firstoftwo
 \else \expandafter \@secondoftwo
 \fi
}%
\providecommand \natexlab [1]{#1}%
\providecommand \enquote  [1]{``#1''}%
\providecommand \bibnamefont  [1]{#1}%
\providecommand \bibfnamefont [1]{#1}%
\providecommand \citenamefont [1]{#1}%
\providecommand \href@noop [0]{\@secondoftwo}%
\providecommand \href [0]{\begingroup \@sanitize@url \@href}%
\providecommand \@href[1]{\@@startlink{#1}\@@href}%
\providecommand \@@href[1]{\endgroup#1\@@endlink}%
\providecommand \@sanitize@url [0]{\catcode `\\12\catcode `\$12\catcode
  `\&12\catcode `\#12\catcode `\^12\catcode `\_12\catcode `\%12\relax}%
\providecommand \@@startlink[1]{}%
\providecommand \@@endlink[0]{}%
\providecommand \url  [0]{\begingroup\@sanitize@url \@url }%
\providecommand \@url [1]{\endgroup\@href {#1}{\urlprefix }}%
\providecommand \urlprefix  [0]{URL }%
\providecommand \Eprint [0]{\href }%
\providecommand \doibase [0]{https://doi.org/}%
\providecommand \selectlanguage [0]{\@gobble}%
\providecommand \bibinfo  [0]{\@secondoftwo}%
\providecommand \bibfield  [0]{\@secondoftwo}%
\providecommand \translation [1]{[#1]}%
\providecommand \BibitemOpen [0]{}%
\providecommand \bibitemStop [0]{}%
\providecommand \bibitemNoStop [0]{.\EOS\space}%
\providecommand \EOS [0]{\spacefactor3000\relax}%
\providecommand \BibitemShut  [1]{\csname bibitem#1\endcsname}%
\let\auto@bib@innerbib\@empty
%</preamble>
\bibitem [{\citenamefont {Datta}\ \emph {et~al.}(2022)\citenamefont {Datta},
  \citenamefont {Ardekani}, \citenamefont {Arratia}, \citenamefont {Beris},
  \citenamefont {Bischofberger}, \citenamefont {McKinley}, \citenamefont
  {Eggers}, \citenamefont {L{\'o}pez-Aguilar}, \citenamefont {Fielding},
  \citenamefont {Frishman} \emph {et~al.}}]{datta2022perspectives}%
  \BibitemOpen
  \bibfield  {author} {\bibinfo {author} {\bibfnamefont {S.~S.}\ \bibnamefont
  {Datta}}, \bibinfo {author} {\bibfnamefont {A.~M.}\ \bibnamefont {Ardekani}},
  \bibinfo {author} {\bibfnamefont {P.~E.}\ \bibnamefont {Arratia}}, \bibinfo
  {author} {\bibfnamefont {A.~N.}\ \bibnamefont {Beris}}, \bibinfo {author}
  {\bibfnamefont {I.}~\bibnamefont {Bischofberger}}, \bibinfo {author}
  {\bibfnamefont {G.~H.}\ \bibnamefont {McKinley}}, \bibinfo {author}
  {\bibfnamefont {J.~G.}\ \bibnamefont {Eggers}}, \bibinfo {author}
  {\bibfnamefont {J.~E.}\ \bibnamefont {L{\'o}pez-Aguilar}}, \bibinfo {author}
  {\bibfnamefont {S.~M.}\ \bibnamefont {Fielding}}, \bibinfo {author}
  {\bibfnamefont {A.}~\bibnamefont {Frishman}}, \emph {et~al.},\ }\bibfield
  {title} {\bibinfo {title} {Perspectives on viscoelastic flow instabilities
  and elastic turbulence},\ }\href@noop {} {\bibfield  {journal} {\bibinfo
  {journal} {Phys. Rev. Fluids}\ }\textbf {\bibinfo {volume} {7}},\ \bibinfo
  {pages} {080701} (\bibinfo {year} {2022})}\BibitemShut {NoStop}%
\bibitem [{\citenamefont {Sanchez}\ \emph {et~al.}(2022)\citenamefont
  {Sanchez}, \citenamefont {Jovanovic}, \citenamefont {Kumar}, \citenamefont
  {Morozov}, \citenamefont {Shankar}, \citenamefont {Subramanian},\ and\
  \citenamefont {Wilson}}]{sanchez22}%
  \BibitemOpen
  \bibfield  {author} {\bibinfo {author} {\bibfnamefont {H.~A.~C.}\
  \bibnamefont {Sanchez}}, \bibinfo {author} {\bibfnamefont {M.~R.}\
  \bibnamefont {Jovanovic}}, \bibinfo {author} {\bibfnamefont {S.}~\bibnamefont
  {Kumar}}, \bibinfo {author} {\bibfnamefont {A.}~\bibnamefont {Morozov}},
  \bibinfo {author} {\bibfnamefont {V.}~\bibnamefont {Shankar}}, \bibinfo
  {author} {\bibfnamefont {G.}~\bibnamefont {Subramanian}},\ and\ \bibinfo
  {author} {\bibfnamefont {H.~J.}\ \bibnamefont {Wilson}},\ }\bibfield  {title}
  {\bibinfo {title} {Understanding viscoelastic flow instabilities:
  {O}ldroyd-{B} and beyond},\ }\href@noop {} {\bibfield  {journal} {\bibinfo
  {journal} {J. Nonnewton. Fluid Mech.}\ }\textbf {\bibinfo {volume} {302}},\
  \bibinfo {pages} {104742} (\bibinfo {year} {2022})}\BibitemShut {NoStop}%
\bibitem [{\citenamefont {Groisman}\ and\ \citenamefont
  {Steinberg}(2000)}]{groisman2000elastic}%
  \BibitemOpen
  \bibfield  {author} {\bibinfo {author} {\bibfnamefont {A.}~\bibnamefont
  {Groisman}}\ and\ \bibinfo {author} {\bibfnamefont {V.}~\bibnamefont
  {Steinberg}},\ }\bibfield  {title} {\bibinfo {title} {Elastic turbulence in a
  polymer solution flow},\ }\href@noop {} {\bibfield  {journal} {\bibinfo
  {journal} {Nature}\ }\textbf {\bibinfo {volume} {405}},\ \bibinfo {pages}
  {53} (\bibinfo {year} {2000})}\BibitemShut {NoStop}%
\bibitem [{\citenamefont {Samanta}\ \emph {et~al.}(2013)\citenamefont
  {Samanta}, \citenamefont {Dubief}, \citenamefont {Holzner}, \citenamefont
  {Sch{\"a}fer}, \citenamefont {Morozov}, \citenamefont {Wagner},\ and\
  \citenamefont {Hof}}]{samanta2013elasto}%
  \BibitemOpen
  \bibfield  {author} {\bibinfo {author} {\bibfnamefont {D.}~\bibnamefont
  {Samanta}}, \bibinfo {author} {\bibfnamefont {Y.}~\bibnamefont {Dubief}},
  \bibinfo {author} {\bibfnamefont {M.}~\bibnamefont {Holzner}}, \bibinfo
  {author} {\bibfnamefont {C.}~\bibnamefont {Sch{\"a}fer}}, \bibinfo {author}
  {\bibfnamefont {A.~N.}\ \bibnamefont {Morozov}}, \bibinfo {author}
  {\bibfnamefont {C.}~\bibnamefont {Wagner}},\ and\ \bibinfo {author}
  {\bibfnamefont {B.}~\bibnamefont {Hof}},\ }\bibfield  {title} {\bibinfo
  {title} {Elasto-inertial turbulence},\ }\href@noop {} {\bibfield  {journal}
  {\bibinfo  {journal} {Proc. Natl. Acad. Sci. U. S. A.}\ }\textbf {\bibinfo
  {volume} {110}},\ \bibinfo {pages} {10557} (\bibinfo {year}
  {2013})}\BibitemShut {NoStop}%
\bibitem [{\citenamefont {Dubief}\ \emph {et~al.}(2023)\citenamefont {Dubief},
  \citenamefont {Terrapon},\ and\ \citenamefont {Hof}}]{dubief2023elasto}%
  \BibitemOpen
  \bibfield  {author} {\bibinfo {author} {\bibfnamefont {Y.}~\bibnamefont
  {Dubief}}, \bibinfo {author} {\bibfnamefont {V.~E.}\ \bibnamefont
  {Terrapon}},\ and\ \bibinfo {author} {\bibfnamefont {B.}~\bibnamefont
  {Hof}},\ }\bibfield  {title} {\bibinfo {title} {Elasto-inertial turbulence},\
  }\href@noop {} {\bibfield  {journal} {\bibinfo  {journal} {Annu. Rev. Fluid
  Mech.}\ }\textbf {\bibinfo {volume} {55}},\ \bibinfo {pages} {675} (\bibinfo
  {year} {2023})}\BibitemShut {NoStop}%
\bibitem [{\citenamefont {Garg}\ \emph {et~al.}(2018)\citenamefont {Garg},
  \citenamefont {Chaudhary}, \citenamefont {Khalid}, \citenamefont {Shankar},\
  and\ \citenamefont {Subramanian}}]{garg2018viscoelastic}%
  \BibitemOpen
  \bibfield  {author} {\bibinfo {author} {\bibfnamefont {P.}~\bibnamefont
  {Garg}}, \bibinfo {author} {\bibfnamefont {I.}~\bibnamefont {Chaudhary}},
  \bibinfo {author} {\bibfnamefont {M.}~\bibnamefont {Khalid}}, \bibinfo
  {author} {\bibfnamefont {V.}~\bibnamefont {Shankar}},\ and\ \bibinfo {author}
  {\bibfnamefont {G.}~\bibnamefont {Subramanian}},\ }\bibfield  {title}
  {\bibinfo {title} {Viscoelastic pipe flow is linearly unstable},\ }\href@noop
  {} {\bibfield  {journal} {\bibinfo  {journal} {Phys. Rev. Lett.}\ }\textbf
  {\bibinfo {volume} {121}},\ \bibinfo {pages} {024502} (\bibinfo {year}
  {2018})}\BibitemShut {NoStop}%
\bibitem [{\citenamefont {Khalid}\ \emph
  {et~al.}(2021{\natexlab{a}})\citenamefont {Khalid}, \citenamefont
  {Chaudhary}, \citenamefont {Garg}, \citenamefont {Shankar},\ and\
  \citenamefont {Subramanian}}]{khalid2021centre}%
  \BibitemOpen
  \bibfield  {author} {\bibinfo {author} {\bibfnamefont {M.}~\bibnamefont
  {Khalid}}, \bibinfo {author} {\bibfnamefont {I.}~\bibnamefont {Chaudhary}},
  \bibinfo {author} {\bibfnamefont {P.}~\bibnamefont {Garg}}, \bibinfo {author}
  {\bibfnamefont {V.}~\bibnamefont {Shankar}},\ and\ \bibinfo {author}
  {\bibfnamefont {G.}~\bibnamefont {Subramanian}},\ }\bibfield  {title}
  {\bibinfo {title} {The centre-mode instability of viscoelastic plane
  {P}oiseuille flow},\ }\href@noop {} {\bibfield  {journal} {\bibinfo
  {journal} {J. Fluid Mech.}\ }\textbf {\bibinfo {volume} {915}} (\bibinfo
  {year} {2021}{\natexlab{a}})}\BibitemShut {NoStop}%
\bibitem [{\citenamefont {Khalid}\ \emph
  {et~al.}(2021{\natexlab{b}})\citenamefont {Khalid}, \citenamefont {Shankar},\
  and\ \citenamefont {Subramanian}}]{khalid2021continuous}%
  \BibitemOpen
  \bibfield  {author} {\bibinfo {author} {\bibfnamefont {M.}~\bibnamefont
  {Khalid}}, \bibinfo {author} {\bibfnamefont {V.}~\bibnamefont {Shankar}},\
  and\ \bibinfo {author} {\bibfnamefont {G.}~\bibnamefont {Subramanian}},\
  }\bibfield  {title} {\bibinfo {title} {Continuous pathway between the
  elasto-inertial and elastic turbulent states in viscoelastic channel flow},\
  }\href@noop {} {\bibfield  {journal} {\bibinfo  {journal} {Phys. Rev. Lett.}\
  }\textbf {\bibinfo {volume} {127}},\ \bibinfo {pages} {134502} (\bibinfo
  {year} {2021}{\natexlab{b}})}\BibitemShut {NoStop}%
\bibitem [{\citenamefont {Page}\ \emph {et~al.}(2020)\citenamefont {Page},
  \citenamefont {Dubief},\ and\ \citenamefont {Kerswell}}]{page2020exact}%
  \BibitemOpen
  \bibfield  {author} {\bibinfo {author} {\bibfnamefont {J.}~\bibnamefont
  {Page}}, \bibinfo {author} {\bibfnamefont {Y.}~\bibnamefont {Dubief}},\ and\
  \bibinfo {author} {\bibfnamefont {R.~R.}\ \bibnamefont {Kerswell}},\
  }\bibfield  {title} {\bibinfo {title} {Exact traveling wave solutions in
  viscoelastic channel flow},\ }\href@noop {} {\bibfield  {journal} {\bibinfo
  {journal} {Phys. Rev. Lett.}\ }\textbf {\bibinfo {volume} {125}},\ \bibinfo
  {pages} {154501} (\bibinfo {year} {2020})}\BibitemShut {NoStop}%
\bibitem [{\citenamefont {Buza}\ \emph {et~al.}(2022)\citenamefont {Buza},
  \citenamefont {Beneitez}, \citenamefont {Page},\ and\ \citenamefont
  {Kerswell}}]{buza2022finite}%
  \BibitemOpen
  \bibfield  {author} {\bibinfo {author} {\bibfnamefont {G.}~\bibnamefont
  {Buza}}, \bibinfo {author} {\bibfnamefont {M.}~\bibnamefont {Beneitez}},
  \bibinfo {author} {\bibfnamefont {J.}~\bibnamefont {Page}},\ and\ \bibinfo
  {author} {\bibfnamefont {R.~R.}\ \bibnamefont {Kerswell}},\ }\bibfield
  {title} {\bibinfo {title} {Finite-amplitude elastic waves in viscoelastic
  channel flow from large to zero reynolds number},\ }\href@noop {} {\bibfield
  {journal} {\bibinfo  {journal} {J. Fluid Mech.}\ }\textbf {\bibinfo {volume}
  {951}},\ \bibinfo {pages} {A3} (\bibinfo {year} {2022})}\BibitemShut
  {NoStop}%
\bibitem [{\citenamefont {Morozov}(2022)}]{morozov2022coherent}%
  \BibitemOpen
  \bibfield  {author} {\bibinfo {author} {\bibfnamefont {A.}~\bibnamefont
  {Morozov}},\ }\bibfield  {title} {\bibinfo {title} {Coherent structures in
  plane channel flow of dilute polymer solutions with vanishing inertia},\
  }\href@noop {} {\bibfield  {journal} {\bibinfo  {journal} {Phys. Rev. Lett.}\
  }\textbf {\bibinfo {volume} {129}},\ \bibinfo {pages} {017801} (\bibinfo
  {year} {2022})}\BibitemShut {NoStop}%
\bibitem [{\citenamefont {Beneitez}\ \emph {et~al.}(2023)\citenamefont
  {Beneitez}, \citenamefont {Page},\ and\ \citenamefont
  {Kerswell}}]{beneitez2023polymer}%
  \BibitemOpen
  \bibfield  {author} {\bibinfo {author} {\bibfnamefont {M.}~\bibnamefont
  {Beneitez}}, \bibinfo {author} {\bibfnamefont {J.}~\bibnamefont {Page}},\
  and\ \bibinfo {author} {\bibfnamefont {R.~R.}\ \bibnamefont {Kerswell}},\
  }\bibfield  {title} {\bibinfo {title} {{Polymer diffusive instability leading
  to elastic turbulence in plane Couette flow}},\ }\href@noop {} {\bibfield
  {journal} {\bibinfo  {journal} {Phys. Rev. Fluids}\ }\textbf {\bibinfo
  {volume} {8}},\ \bibinfo {pages} {L101901} (\bibinfo {year}
  {2023})}\BibitemShut {NoStop}%
\bibitem [{\citenamefont {Lellep}\ \emph {et~al.}(2024)\citenamefont {Lellep},
  \citenamefont {Linkmann},\ and\ \citenamefont {Morozov}}]{lellep2024purely}%
  \BibitemOpen
  \bibfield  {author} {\bibinfo {author} {\bibfnamefont {M.}~\bibnamefont
  {Lellep}}, \bibinfo {author} {\bibfnamefont {M.}~\bibnamefont {Linkmann}},\
  and\ \bibinfo {author} {\bibfnamefont {A.}~\bibnamefont {Morozov}},\
  }\bibfield  {title} {\bibinfo {title} {Purely elastic turbulence in
  pressure-driven channel flows},\ }\href@noop {} {\bibfield  {journal}
  {\bibinfo  {journal} {Proc. Natl. Acad. Sci. U.S.A.}\ }\textbf {\bibinfo
  {volume} {121}},\ \bibinfo {pages} {e2318851121} (\bibinfo {year}
  {2024})}\BibitemShut {NoStop}%
\bibitem [{\citenamefont {Larson}\ \emph {et~al.}(1990)\citenamefont {Larson},
  \citenamefont {Shaqfeh},\ and\ \citenamefont {Muller}}]{larson1990purely}%
  \BibitemOpen
  \bibfield  {author} {\bibinfo {author} {\bibfnamefont {R.~G.}\ \bibnamefont
  {Larson}}, \bibinfo {author} {\bibfnamefont {E.~S.}\ \bibnamefont
  {Shaqfeh}},\ and\ \bibinfo {author} {\bibfnamefont {S.~J.}\ \bibnamefont
  {Muller}},\ }\bibfield  {title} {\bibinfo {title} {A purely elastic
  instability in {T}aylor--{C}ouette flow},\ }\href@noop {} {\bibfield
  {journal} {\bibinfo  {journal} {J. Fluid Mech.}\ }\textbf {\bibinfo {volume}
  {218}},\ \bibinfo {pages} {573} (\bibinfo {year} {1990})}\BibitemShut
  {NoStop}%
\bibitem [{\citenamefont {Pakdel}\ and\ \citenamefont
  {McKinley}(1996)}]{pakdel1996elastic}%
  \BibitemOpen
  \bibfield  {author} {\bibinfo {author} {\bibfnamefont {P.}~\bibnamefont
  {Pakdel}}\ and\ \bibinfo {author} {\bibfnamefont {G.~H.}\ \bibnamefont
  {McKinley}},\ }\bibfield  {title} {\bibinfo {title} {Elastic instability and
  curved streamlines},\ }\href@noop {} {\bibfield  {journal} {\bibinfo
  {journal} {Phys. Rev. Lett.}\ }\textbf {\bibinfo {volume} {77}},\ \bibinfo
  {pages} {2459} (\bibinfo {year} {1996})}\BibitemShut {NoStop}%
\bibitem [{\citenamefont {Pan}\ \emph {et~al.}(2013)\citenamefont {Pan},
  \citenamefont {Morozov}, \citenamefont {Wagner},\ and\ \citenamefont
  {Arratia}}]{pan2013nonlinear}%
  \BibitemOpen
  \bibfield  {author} {\bibinfo {author} {\bibfnamefont {L.}~\bibnamefont
  {Pan}}, \bibinfo {author} {\bibfnamefont {A.}~\bibnamefont {Morozov}},
  \bibinfo {author} {\bibfnamefont {C.}~\bibnamefont {Wagner}},\ and\ \bibinfo
  {author} {\bibfnamefont {P.}~\bibnamefont {Arratia}},\ }\bibfield  {title}
  {\bibinfo {title} {Nonlinear elastic instability in channel flows at low
  {R}eynolds numbers},\ }\href@noop {} {\bibfield  {journal} {\bibinfo
  {journal} {Phys. Rev. Lett.}\ }\textbf {\bibinfo {volume} {110}},\ \bibinfo
  {pages} {174502} (\bibinfo {year} {2013})}\BibitemShut {NoStop}%
\bibitem [{\citenamefont {Sid}\ \emph {et~al.}(2018)\citenamefont {Sid},
  \citenamefont {Terrapon},\ and\ \citenamefont {Dubief}}]{sid2018two}%
  \BibitemOpen
  \bibfield  {author} {\bibinfo {author} {\bibfnamefont {S.}~\bibnamefont
  {Sid}}, \bibinfo {author} {\bibfnamefont {V.}~\bibnamefont {Terrapon}},\ and\
  \bibinfo {author} {\bibfnamefont {Y.}~\bibnamefont {Dubief}},\ }\bibfield
  {title} {\bibinfo {title} {Two-dimensional dynamics of elasto-inertial
  turbulence and its role in polymer drag reduction},\ }\href@noop {}
  {\bibfield  {journal} {\bibinfo  {journal} {Phys. Rev. Fluids}\ }\textbf
  {\bibinfo {volume} {3}},\ \bibinfo {pages} {011301} (\bibinfo {year}
  {2018})}\BibitemShut {NoStop}%
\bibitem [{\citenamefont {Zhang}\ \emph {et~al.}(2013)\citenamefont {Zhang},
  \citenamefont {Lashgari}, \citenamefont {Zaki},\ and\ \citenamefont
  {Brandt}}]{zhang2013linear}%
  \BibitemOpen
  \bibfield  {author} {\bibinfo {author} {\bibfnamefont {M.}~\bibnamefont
  {Zhang}}, \bibinfo {author} {\bibfnamefont {I.}~\bibnamefont {Lashgari}},
  \bibinfo {author} {\bibfnamefont {T.~A.}\ \bibnamefont {Zaki}},\ and\
  \bibinfo {author} {\bibfnamefont {L.}~\bibnamefont {Brandt}},\ }\bibfield
  {title} {\bibinfo {title} {Linear stability analysis of channel flow of
  viscoelastic oldroyd-b and fene-p fluids},\ }\href@noop {} {\bibfield
  {journal} {\bibinfo  {journal} {J. Fluid Mech.}\ }\textbf {\bibinfo {volume}
  {737}},\ \bibinfo {pages} {249} (\bibinfo {year} {2013})}\BibitemShut
  {NoStop}%
\bibitem [{\citenamefont {Lellep}\ \emph {et~al.}(2023)\citenamefont {Lellep},
  \citenamefont {Linkmann},\ and\ \citenamefont {Morozov}}]{lellep2023linear}%
  \BibitemOpen
  \bibfield  {author} {\bibinfo {author} {\bibfnamefont {M.}~\bibnamefont
  {Lellep}}, \bibinfo {author} {\bibfnamefont {M.}~\bibnamefont {Linkmann}},\
  and\ \bibinfo {author} {\bibfnamefont {A.}~\bibnamefont {Morozov}},\
  }\bibfield  {title} {\bibinfo {title} {Linear stability analysis of purely
  elastic travelling-wave solutions in pressure-driven channel flows},\
  }\href@noop {} {\bibfield  {journal} {\bibinfo  {journal} {J. Fluid Mech.}\
  }\textbf {\bibinfo {volume} {959}},\ \bibinfo {pages} {R1} (\bibinfo {year}
  {2023})}\BibitemShut {NoStop}%
\bibitem [{\citenamefont {Shekar}\ \emph {et~al.}(2019)\citenamefont {Shekar},
  \citenamefont {McMullen}, \citenamefont {Wang}, \citenamefont {McKeon},\ and\
  \citenamefont {Graham}}]{shekar2019critical}%
  \BibitemOpen
  \bibfield  {author} {\bibinfo {author} {\bibfnamefont {A.}~\bibnamefont
  {Shekar}}, \bibinfo {author} {\bibfnamefont {R.~M.}\ \bibnamefont
  {McMullen}}, \bibinfo {author} {\bibfnamefont {S.-N.}\ \bibnamefont {Wang}},
  \bibinfo {author} {\bibfnamefont {B.~J.}\ \bibnamefont {McKeon}},\ and\
  \bibinfo {author} {\bibfnamefont {M.~D.}\ \bibnamefont {Graham}},\ }\bibfield
   {title} {\bibinfo {title} {Critical-layer structures and mechanisms in
  elastoinertial turbulence},\ }\href@noop {} {\bibfield  {journal} {\bibinfo
  {journal} {Phys. Rev. Lett.}\ }\textbf {\bibinfo {volume} {122}},\ \bibinfo
  {pages} {124503} (\bibinfo {year} {2019})}\BibitemShut {NoStop}%
\bibitem [{\citenamefont {Shekar}\ \emph {et~al.}(2021)\citenamefont {Shekar},
  \citenamefont {McMullen}, \citenamefont {McKeon},\ and\ \citenamefont
  {Graham}}]{shekar2021tollmien}%
  \BibitemOpen
  \bibfield  {author} {\bibinfo {author} {\bibfnamefont {A.}~\bibnamefont
  {Shekar}}, \bibinfo {author} {\bibfnamefont {R.~M.}\ \bibnamefont
  {McMullen}}, \bibinfo {author} {\bibfnamefont {B.~J.}\ \bibnamefont
  {McKeon}},\ and\ \bibinfo {author} {\bibfnamefont {M.~D.}\ \bibnamefont
  {Graham}},\ }\bibfield  {title} {\bibinfo {title} {Tollmien-schlichting route
  to elastoinertial turbulence in channel flow},\ }\href@noop {} {\bibfield
  {journal} {\bibinfo  {journal} {Phys. Rev. Fluids}\ }\textbf {\bibinfo
  {volume} {6}},\ \bibinfo {pages} {093301} (\bibinfo {year}
  {2021})}\BibitemShut {NoStop}%
\bibitem [{\citenamefont {Beneitez}\ \emph {et~al.}(2024)\citenamefont
  {Beneitez}, \citenamefont {Page}, \citenamefont {Dubief},\ and\ \citenamefont
  {Kerswell}}]{beneitez2024multistability}%
  \BibitemOpen
  \bibfield  {author} {\bibinfo {author} {\bibfnamefont {M.}~\bibnamefont
  {Beneitez}}, \bibinfo {author} {\bibfnamefont {J.}~\bibnamefont {Page}},
  \bibinfo {author} {\bibfnamefont {Y.}~\bibnamefont {Dubief}},\ and\ \bibinfo
  {author} {\bibfnamefont {R.~R.}\ \bibnamefont {Kerswell}},\ }\bibfield
  {title} {\bibinfo {title} {Multistability of elasto-inertial two-dimensional
  channel flow},\ }\href@noop {} {\bibfield  {journal} {\bibinfo  {journal} {J.
  Fluid Mech.}\ }\textbf {\bibinfo {volume} {981}},\ \bibinfo {pages} {A30}
  (\bibinfo {year} {2024})}\BibitemShut {NoStop}%
\bibitem [{\citenamefont {Couchman}\ \emph {et~al.}(2024)\citenamefont
  {Couchman}, \citenamefont {Beneitez}, \citenamefont {Page},\ and\
  \citenamefont {Kerswell}}]{couchman2024inertial}%
  \BibitemOpen
  \bibfield  {author} {\bibinfo {author} {\bibfnamefont {M.~M.}\ \bibnamefont
  {Couchman}}, \bibinfo {author} {\bibfnamefont {M.}~\bibnamefont {Beneitez}},
  \bibinfo {author} {\bibfnamefont {J.}~\bibnamefont {Page}},\ and\ \bibinfo
  {author} {\bibfnamefont {R.~R.}\ \bibnamefont {Kerswell}},\ }\bibfield
  {title} {\bibinfo {title} {Inertial enhancement of the polymer diffusive
  instability},\ }\href@noop {} {\bibfield  {journal} {\bibinfo  {journal} {J.
  Fluid Mech.}\ }\textbf {\bibinfo {volume} {981}},\ \bibinfo {pages} {A2}
  (\bibinfo {year} {2024})}\BibitemShut {NoStop}%
\bibitem [{\citenamefont {Lewy}\ and\ \citenamefont {Kerswell}(2024)}]{lewy24}%
  \BibitemOpen
  \bibfield  {author} {\bibinfo {author} {\bibfnamefont {T.}~\bibnamefont
  {Lewy}}\ and\ \bibinfo {author} {\bibfnamefont {R.~R.}\ \bibnamefont
  {Kerswell}},\ }\bibfield  {title} {\bibinfo {title} {The polymer diffusive
  instability in highly concentrated polymeric fluids},\ }\href@noop {}
  {\bibfield  {journal} {\bibinfo  {journal} {Journal of Non-Newtonian Fluid
  Mechanics}\ }\textbf {\bibinfo {volume} {326}} (\bibinfo {year}
  {2024})}\BibitemShut {NoStop}%
\bibitem [{\citenamefont {Page}\ and\ \citenamefont {Zaki}(2022)}]{Page2022}%
  \BibitemOpen
  \bibfield  {author} {\bibinfo {author} {\bibfnamefont {J.}~\bibnamefont
  {Page}}\ and\ \bibinfo {author} {\bibfnamefont {T.~A.}\ \bibnamefont
  {Zaki}},\ }\bibfield  {title} {\bibinfo {title} {Vorticity amplification in
  wavy viscoelastic channel flow},\ }\href@noop {} {\bibfield  {journal}
  {\bibinfo  {journal} {J. Fluid Mech.}\ }\textbf {\bibinfo {volume} {949}}
  (\bibinfo {year} {2022})}\BibitemShut {NoStop}%
\bibitem [{\citenamefont {Sureshkumar}\ \emph {et~al.}(1997)\citenamefont
  {Sureshkumar}, \citenamefont {Beris},\ and\ \citenamefont
  {Handler}}]{sureshkumar1997direct}%
  \BibitemOpen
  \bibfield  {author} {\bibinfo {author} {\bibfnamefont {R.}~\bibnamefont
  {Sureshkumar}}, \bibinfo {author} {\bibfnamefont {A.~N.}\ \bibnamefont
  {Beris}},\ and\ \bibinfo {author} {\bibfnamefont {R.~A.}\ \bibnamefont
  {Handler}},\ }\bibfield  {title} {\bibinfo {title} {Direct numerical
  simulation of the turbulent channel flow of a polymer solution},\ }\href@noop
  {} {\bibfield  {journal} {\bibinfo  {journal} {Phys. Fluids}\ }\textbf
  {\bibinfo {volume} {9}},\ \bibinfo {pages} {743} (\bibinfo {year}
  {1997})}\BibitemShut {NoStop}%
\bibitem [{\citenamefont {Hiemenz}\ and\ \citenamefont
  {Lodge}(2007)}]{hiemenz2007polymer}%
  \BibitemOpen
  \bibfield  {author} {\bibinfo {author} {\bibfnamefont {P.~C.}\ \bibnamefont
  {Hiemenz}}\ and\ \bibinfo {author} {\bibfnamefont {T.~P.}\ \bibnamefont
  {Lodge}},\ }\href@noop {} {\emph {\bibinfo {title} {{Polymer Chemistry}}}}\
  (\bibinfo  {publisher} {CRC press},\ \bibinfo {year} {2007})\BibitemShut
  {NoStop}%
\bibitem [{\citenamefont {Ruelle}\ and\ \citenamefont
  {Takens}(1971)}]{ruelle1971nature}%
  \BibitemOpen
  \bibfield  {author} {\bibinfo {author} {\bibfnamefont {D.}~\bibnamefont
  {Ruelle}}\ and\ \bibinfo {author} {\bibfnamefont {F.}~\bibnamefont
  {Takens}},\ }\bibfield  {title} {\bibinfo {title} {On the nature of
  turbulence},\ }\href@noop {} {\bibfield  {journal} {\bibinfo  {journal} {Les
  rencontres physiciens-math{\'e}maticiens de Strasbourg-RCP25}\ }\textbf
  {\bibinfo {volume} {12}},\ \bibinfo {pages} {1} (\bibinfo {year}
  {1971})}\BibitemShut {NoStop}%
\bibitem [{\citenamefont {Berti}\ \emph {et~al.}(2008)\citenamefont {Berti},
  \citenamefont {Bistagnino}, \citenamefont {Boffetta}, \citenamefont
  {Celani},\ and\ \citenamefont {Musacchio}}]{berti2008}%
  \BibitemOpen
  \bibfield  {author} {\bibinfo {author} {\bibfnamefont {S.}~\bibnamefont
  {Berti}}, \bibinfo {author} {\bibfnamefont {A.}~\bibnamefont {Bistagnino}},
  \bibinfo {author} {\bibfnamefont {G.}~\bibnamefont {Boffetta}}, \bibinfo
  {author} {\bibfnamefont {A.}~\bibnamefont {Celani}},\ and\ \bibinfo {author}
  {\bibfnamefont {S.}~\bibnamefont {Musacchio}},\ }\bibfield  {title} {\bibinfo
  {title} {Two-dimensional elastic turbulence},\ }\href@noop {} {\bibfield
  {journal} {\bibinfo  {journal} {Physical Review E—Statistical, Nonlinear,
  and Soft Matter Physics}\ }\textbf {\bibinfo {volume} {77}},\ \bibinfo
  {pages} {055306} (\bibinfo {year} {2008})}\BibitemShut {NoStop}%
\bibitem [{\citenamefont {Berti}\ and\ \citenamefont
  {Boffetta}(2010)}]{berti2010}%
  \BibitemOpen
  \bibfield  {author} {\bibinfo {author} {\bibfnamefont {S.}~\bibnamefont
  {Berti}}\ and\ \bibinfo {author} {\bibfnamefont {G.}~\bibnamefont
  {Boffetta}},\ }\bibfield  {title} {\bibinfo {title} {Elastic waves and
  transition to elastic turbulence in a two-dimensional viscoelastic kolmogorov
  flow},\ }\href {https://doi.org/10.1103/PhysRevE.82.036314} {\bibfield
  {journal} {\bibinfo  {journal} {Phys. Rev. E}\ }\textbf {\bibinfo {volume}
  {82}},\ \bibinfo {pages} {036314} (\bibinfo {year} {2010})}\BibitemShut
  {NoStop}%
\bibitem [{\citenamefont {Rota}\ \emph {et~al.}(2023)\citenamefont {Rota},
  \citenamefont {Amor}, \citenamefont {Clainche},\ and\ \citenamefont
  {Rosti}}]{rota2023elastic}%
  \BibitemOpen
  \bibfield  {author} {\bibinfo {author} {\bibfnamefont {G.~F.}\ \bibnamefont
  {Rota}}, \bibinfo {author} {\bibfnamefont {C.}~\bibnamefont {Amor}}, \bibinfo
  {author} {\bibfnamefont {S.~L.}\ \bibnamefont {Clainche}},\ and\ \bibinfo
  {author} {\bibfnamefont {M.~E.}\ \bibnamefont {Rosti}},\ }\bibfield  {title}
  {\bibinfo {title} {Elastic turbulence in planar channel flows--turbulence
  with no drag and enhanced mixing},\ }\href@noop {} {\bibfield  {journal}
  {\bibinfo  {journal} {arXiv preprint arXiv:2310.05340}\ } (\bibinfo {year}
  {2023})}\BibitemShut {NoStop}%
\end{thebibliography}%


\providecommand{\noopsort}[1]{}\providecommand{\singleletter}[1]{#1}%
\begin{thebibliography}{3}%
\makeatletter
\providecommand \@ifxundefined [1]{%
 \@ifx{#1\undefined}
}%
\providecommand \@ifnum [1]{%
 \ifnum #1\expandafter \@firstoftwo
 \else \expandafter \@secondoftwo
 \fi
}%
\providecommand \@ifx [1]{%
 \ifx #1\expandafter \@firstoftwo
 \else \expandafter \@secondoftwo
 \fi
}%
\providecommand \natexlab [1]{#1}%
\providecommand \enquote  [1]{``#1''}%
\providecommand \bibnamefont  [1]{#1}%
\providecommand \bibfnamefont [1]{#1}%
\providecommand \citenamefont [1]{#1}%
\providecommand \href@noop [0]{\@secondoftwo}%
\providecommand \href [0]{\begingroup \@sanitize@url \@href}%
\providecommand \@href[1]{\@@startlink{#1}\@@href}%
\providecommand \@@href[1]{\endgroup#1\@@endlink}%
\providecommand \@sanitize@url [0]{\catcode `\\12\catcode `\$12\catcode
  `\&12\catcode `\#12\catcode `\^12\catcode `\_12\catcode `\%12\relax}%
\providecommand \@@startlink[1]{}%
\providecommand \@@endlink[0]{}%
\providecommand \url  [0]{\begingroup\@sanitize@url \@url }%
\providecommand \@url [1]{\endgroup\@href {#1}{\urlprefix }}%
\providecommand \urlprefix  [0]{URL }%
\providecommand \Eprint [0]{\href }%
\providecommand \doibase [0]{https://doi.org/}%
\providecommand \selectlanguage [0]{\@gobble}%
\providecommand \bibinfo  [0]{\@secondoftwo}%
\providecommand \bibfield  [0]{\@secondoftwo}%
\providecommand \translation [1]{[#1]}%
\providecommand \BibitemOpen [0]{}%
\providecommand \bibitemStop [0]{}%
\providecommand \bibitemNoStop [0]{.\EOS\space}%
\providecommand \EOS [0]{\spacefactor3000\relax}%
\providecommand \BibitemShut  [1]{\csname bibitem#1\endcsname}%
\let\auto@bib@innerbib\@empty
%</preamble>
\bibitem [{\citenamefont {Burns}\ \emph {et~al.}(2020)\citenamefont {Burns},
  \citenamefont {Vasil}, \citenamefont {Oishi}, \citenamefont {Lecoanet},\ and\
  \citenamefont {Brown}}]{burns2020dedalus}%
  \BibitemOpen
  \bibfield  {author} {\bibinfo {author} {\bibfnamefont {K.~J.}\ \bibnamefont
  {Burns}}, \bibinfo {author} {\bibfnamefont {G.~M.}\ \bibnamefont {Vasil}},
  \bibinfo {author} {\bibfnamefont {J.~S.}\ \bibnamefont {Oishi}}, \bibinfo
  {author} {\bibfnamefont {D.}~\bibnamefont {Lecoanet}},\ and\ \bibinfo
  {author} {\bibfnamefont {B.~P.}\ \bibnamefont {Brown}},\ }\bibfield  {title}
  {\bibinfo {title} {Dedalus: A flexible framework for numerical simulations
  with spectral methods},\ }\href@noop {} {\bibfield  {journal} {\bibinfo
  {journal} {Phys. Rev. Research}\ }\textbf {\bibinfo {volume} {2}},\ \bibinfo
  {pages} {023068} (\bibinfo {year} {2020})}\BibitemShut {NoStop}%
\bibitem [{\citenamefont {Virtanen}\ \emph {et~al.}(2020)\citenamefont
  {Virtanen}, \citenamefont {Gommers}, \citenamefont {Oliphant}, \citenamefont
  {Haberland}, \citenamefont {Reddy}, \citenamefont {Cournapeau}, \citenamefont
  {Burovski}, \citenamefont {Peterson}, \citenamefont {Weckesser},
  \citenamefont {Bright}, \citenamefont {{van der Walt}}, \citenamefont
  {Brett}, \citenamefont {Wilson}, \citenamefont {Millman}, \citenamefont
  {Mayorov}, \citenamefont {Nelson}, \citenamefont {Jones}, \citenamefont
  {Kern}, \citenamefont {Larson}, \citenamefont {Carey}, \citenamefont {Polat},
  \citenamefont {Feng}, \citenamefont {Moore}, \citenamefont {{VanderPlas}},
  \citenamefont {Laxalde}, \citenamefont {Perktold}, \citenamefont {Cimrman},
  \citenamefont {Henriksen}, \citenamefont {Quintero}, \citenamefont {Harris},
  \citenamefont {Archibald}, \citenamefont {Ribeiro}, \citenamefont
  {Pedregosa}, \citenamefont {{van Mulbregt}},\ and\ \citenamefont {{SciPy 1.0
  Contributors}}}]{2020SciPy-NMeth}%
  \BibitemOpen
  \bibfield  {author} {\bibinfo {author} {\bibfnamefont {P.}~\bibnamefont
  {Virtanen}}, \bibinfo {author} {\bibfnamefont {R.}~\bibnamefont {Gommers}},
  \bibinfo {author} {\bibfnamefont {T.~E.}\ \bibnamefont {Oliphant}}, \bibinfo
  {author} {\bibfnamefont {M.}~\bibnamefont {Haberland}}, \bibinfo {author}
  {\bibfnamefont {T.}~\bibnamefont {Reddy}}, \bibinfo {author} {\bibfnamefont
  {D.}~\bibnamefont {Cournapeau}}, \bibinfo {author} {\bibfnamefont
  {E.}~\bibnamefont {Burovski}}, \bibinfo {author} {\bibfnamefont
  {P.}~\bibnamefont {Peterson}}, \bibinfo {author} {\bibfnamefont
  {W.}~\bibnamefont {Weckesser}}, \bibinfo {author} {\bibfnamefont
  {J.}~\bibnamefont {Bright}}, \bibinfo {author} {\bibfnamefont {S.~J.}\
  \bibnamefont {{van der Walt}}}, \bibinfo {author} {\bibfnamefont
  {M.}~\bibnamefont {Brett}}, \bibinfo {author} {\bibfnamefont
  {J.}~\bibnamefont {Wilson}}, \bibinfo {author} {\bibfnamefont {K.~J.}\
  \bibnamefont {Millman}}, \bibinfo {author} {\bibfnamefont {N.}~\bibnamefont
  {Mayorov}}, \bibinfo {author} {\bibfnamefont {A.~R.~J.}\ \bibnamefont
  {Nelson}}, \bibinfo {author} {\bibfnamefont {E.}~\bibnamefont {Jones}},
  \bibinfo {author} {\bibfnamefont {R.}~\bibnamefont {Kern}}, \bibinfo {author}
  {\bibfnamefont {E.}~\bibnamefont {Larson}}, \bibinfo {author} {\bibfnamefont
  {C.~J.}\ \bibnamefont {Carey}}, \bibinfo {author} {\bibfnamefont
  {{\.I}.}~\bibnamefont {Polat}}, \bibinfo {author} {\bibfnamefont
  {Y.}~\bibnamefont {Feng}}, \bibinfo {author} {\bibfnamefont {E.~W.}\
  \bibnamefont {Moore}}, \bibinfo {author} {\bibfnamefont {J.}~\bibnamefont
  {{VanderPlas}}}, \bibinfo {author} {\bibfnamefont {D.}~\bibnamefont
  {Laxalde}}, \bibinfo {author} {\bibfnamefont {J.}~\bibnamefont {Perktold}},
  \bibinfo {author} {\bibfnamefont {R.}~\bibnamefont {Cimrman}}, \bibinfo
  {author} {\bibfnamefont {I.}~\bibnamefont {Henriksen}}, \bibinfo {author}
  {\bibfnamefont {E.~A.}\ \bibnamefont {Quintero}}, \bibinfo {author}
  {\bibfnamefont {C.~R.}\ \bibnamefont {Harris}}, \bibinfo {author}
  {\bibfnamefont {A.~M.}\ \bibnamefont {Archibald}}, \bibinfo {author}
  {\bibfnamefont {A.~H.}\ \bibnamefont {Ribeiro}}, \bibinfo {author}
  {\bibfnamefont {F.}~\bibnamefont {Pedregosa}}, \bibinfo {author}
  {\bibfnamefont {P.}~\bibnamefont {{van Mulbregt}}},\ and\ \bibinfo {author}
  {\bibnamefont {{SciPy 1.0 Contributors}}},\ }\bibfield  {title} {\bibinfo
  {title} {{{SciPy} 1.0: Fundamental Algorithms for Scientific Computing in
  Python}},\ }\href {https://doi.org/10.1038/s41592-019-0686-2} {\bibfield
  {journal} {\bibinfo  {journal} {Nature Methods}\ }\textbf {\bibinfo {volume}
  {17}},\ \bibinfo {pages} {261} (\bibinfo {year} {2020})}\BibitemShut
  {NoStop}%
\bibitem [{\citenamefont {Lehoucq}\ \emph {et~al.}(1998)\citenamefont
  {Lehoucq}, \citenamefont {Sorensen},\ and\ \citenamefont
  {Yang}}]{lehoucq1998arpack}%
  \BibitemOpen
  \bibfield  {author} {\bibinfo {author} {\bibfnamefont {R.~B.}\ \bibnamefont
  {Lehoucq}}, \bibinfo {author} {\bibfnamefont {D.~C.}\ \bibnamefont
  {Sorensen}},\ and\ \bibinfo {author} {\bibfnamefont {C.}~\bibnamefont
  {Yang}},\ }\href@noop {} {\emph {\bibinfo {title} {ARPACK users' guide:
  solution of large-scale eigenvalue problems with implicitly restarted Arnoldi
  methods}}}\ (\bibinfo  {publisher} {SIAM},\ \bibinfo {year}
  {1998})\BibitemShut {NoStop}%
\end{thebibliography}%

\end{document}


\baselineskip24pt


\maketitle 

\paragraph*{This PDF includes:}

Materials and Methods

\section*{Materials and Methods}
\paragraph{Linear stability analysis.} The Navier--Stokes equations with the FENE-P viscoelastic model in the Stokes flow limit are
\begin{subequations}
\begin{align}
    \nabla p &= \beta\Delta \mathbf{u}+(1-\beta)\nabla \cdot \mathbf{T}(\mathbf{C}),\\
    \partial_t \mathbf{C}+ (\mathbf{u}\cdot \nabla ) \mathbf{C} + \mathbf{T}(\mathbf{C}) &= \mathbf{C}\cdot \nabla \mathbf{u} + (\nabla \mathbf{u})^T\cdot \mathbf{C}+ \varepsilon\Delta \mathbf{C}, \label{eq:Ceq_Re0} \\
    \nabla \cdot \mathbf{u} &= 0,
\end{align}\label{eq:stokesFlow}
\end{subequations}
with
\begin{align}
    \mathbf{T}(\mathbf{C}) := \frac{1}{\textit{W}}\left(f(\text{tr}\ \mathbf{C})\mathbf{C}-\mathbf{I}\right), \quad \text{with}\ f(s) = \left(1-\frac{s-3}{L_{\text{max}}^2}\right)^{-1}  \label{eq:constModel}
\end{align}
where $\mathbf{u}$ denotes the velocity field, $p$ the pressure and $\mathbf{C}$  is the positive-definite conformation tensor which represents the local phase average of the product of the end-to-end vector of each polymeric molecule. $\beta:=\nu_s/\nu$ denotes the viscosity ratio, where $\nu_s$ and $\nu_p:=\nu-\nu_s$ are the solvent and polymer kinematic viscosity respectively. $L_{\text{max}}$ is the maximum extensibility of the polymer (with the limit $L_{\text{max}}\to\infty$ in \eqref{eq:constModel} recovering the Oldroyd-B model). The non-dimensional is the Weissenberg number $W:=U\lambda/L$, for some characteristic velocity $U$, lengthscale $L$ and polymer relaxation time $\lambda$. The last term in the polymer equation accounts for the polymer diffusion. 

We consider a two-dimensional configuration with a streamwise $x$ direction and a wall-normal $y$ direction. The equations are normalised with the characteristic lengthscale of half the distance between the walls, $h$. The case under consideration is plane Couette flow (PCF) and the characteristic speed $U$ is half the speed difference between the upper and lower wall. 
Adding a small perturbation, $\phi^*$, to the base state $\Phi$ so that the total  field is $\phi=\Phi+\phi^{*}$, the linearised equations are
\begin{subequations}
\begin{align}
    \partial_x p^* &= \beta\left(\partial_{xx}u^* + \partial_{yy}u^* \right) + (1-\beta)\left(\partial_x \tau^*_{xx} + \partial_{y}\tau^*_{xy} \right),\\
    \partial_y p^* &= \beta\left(\partial_{xx}v^* + \partial_{yy}v^* \right) + (1-\beta)\left(\partial_x \tau^*_{xy} + \partial_{y}\tau^*_{yy} \right),\\
    \partial_t c^*_{xx}+U\partial_x c^*_{xx}+C_{xx}' v^{*} +\tau^*_{xx}&=2\left(C_{xx}\partial_x u^*+U' c^*_{xy}+C_{xy}\partial_y u^{*}\right) + \varepsilon\left(\partial_{xx}c^*_{xx}+\partial_{yy}c^*_{xx}\right),\\
    \partial_t c^*_{yy}+U\partial_x c^*_{yy}+C_{yy}' v^{*} +\tau^*_{yy}&=2\left(C_{xy}\partial_x+C_{yy}\partial_y v^{*}\right) + \varepsilon\left(\partial_{xx}c^*_{yy}+\partial_{yy}c^*_{yy}\right),\\
    \partial_t c^*_{zz}+U\partial_x c^*_{zz}+C_{zz}' v^{*} +\tau^*_{zz}&= \varepsilon\left(\partial_{xx}c^*_{zz}+\partial_{yy}c^*_{zz}\right),\\
    \partial_t c^*_{xy}+U\partial_x c^*_{xy}+C_{xy}' v^{*} +\tau^*_{xy}&=\left(C_xx\partial_x v^{*}+U'c^*_{yy}+C_{yy}\partial_y u^{*}\right) + \varepsilon\left(\partial_{xx}c^*_{xy}+\partial_{yy}c^*_{xy}\right),\\
    \partial_x u^{*}+\partial_y v^{*} = 0,
\end{align}
\end{subequations}
where 
\begin{equation}
    \tau^*_{ij} = \frac{1}{W}\left(\text{tr}(\mathbf{c}^{*})\frac{f^2(\text{tr}\ \mathbf{C})}{L_{\text{max}}^2}C_{ij}+f(\text{tr}\ \mathbf{C}) c^*_{ij}\right),
\end{equation}
and
\begin{subequations}
\begin{align}
    \partial_x \tau_{ij}^*&=\frac{1}{W}\left(\partial_x\text{tr}\ \mathbf{c}\frac{f^2(\text{tr}\ \mathbf{C})}{L_{\text{max}}^2}C_{ij}+f(\text{tr}\ \mathbf{C})\partial_x c^*_{ij}\right),\\
    \partial_y \tau_{ij}^*&=\frac{1}{W}\left(\partial_y\text{tr}\ \mathbf{c}\frac{f^2(\text{tr}\ \mathbf{C})}{L_{\text{max}}^2}C_{ij}+\text{tr}\ \mathbf{C}\partial_y \left[\frac{f^2(\text{tr}\ \mathbf{C})}{L_{\text{max}}^2}C_{ij}\right] + c^*_{ij}\partial_y f(\text{tr}\ \mathbf{C}) +  f(\text{tr}\ \mathbf{C})\partial_y c^*_{ij}\right).
\end{align}
\end{subequations}
(primed quantities denote base flow derivatives with respect to the wall-normal direction).
We consider the exponential growth of perturbations and apply a Fourier transform in the homogeneous $x$ direction  by introducing the ansatz $\phi^* = \hat{\phi}(y)e^{i k_x(x-c t)}$, where $k_x$ is the the streamwise wavenumber and $c=c_r+ic_i$ the complex phase speed of the perturbations, with $c_i>0$ for linear instability.

\paragraph{Numerical Method.} The direct numerical simulations (DNS) of eqs. (\ref{eq:stokesFlow}) are performed using the numerical codebase Dedalus \textit{(24)}. We consider a computational domain of size $[L_x,L_y,L_z]=[2\pi,2,2\pi]$ in units of $h$. The quantities $\mathbf{C}$ and $\mathbf{u}$ are expanded in $N_x$ and $N_z$ Fourier modes in the streamwise ($x$) and spanwise ($z$) directions respectively. An expansion in $N_y$ Chebyshev modes is employed in the wall-normal ($y$) direction. Time integration is performed with a 4th-order exponential Runge-Kutta (RK443) scheme with fixed time step $\Delta t=5\times 10^{-3}$. The two-dimensional simulations were performed with a resolution $[N_x,N_y]=[128,256]$ and the three-dimensional ones used $[N_x,N_y,N_z]=[64,128,32]$. The consistency of the results was checked with double the resolution in both cases, obtaining the same dynamical behaviour. The Dedalus codebase has been extensively used in viscoelastic flow simulation \textit{(24,18,19)}. The two-dimensional simulations were initialised by adding random noise to the undisturbed laminar state. We tested the three-dimensional stability of the two-dimensional state by adding small am plitude random perturbation in the spanwise component of the velocity $w$.

%\RK{ok  looks good...am off home for lunch - just given my 12 noon lecture.}

%\MB{Very nice. Thank you very much for going through everything here Rich. I'm happy with the changes and will just wait for Jacob to check everything.}

%\JP{Hi all. I'm happy!}

\preprint{APS/123-QED}

% \title{A generic pathway to viscoelastic turbulence in polymer flows}% Force line breaks with \\ % to polymeric turbulence // to non-Newtonian turbulence // 
\title{Supplementary Material \\ Transition route to elaststic and elasto-inertial turbulence in polymer channel flows}

\author{M. Beneitez}
%\email{mb2467@cam.ac.uk}
\affiliation{
 DAMTP, Centre for Mathematical Sciences, Wilberforce Road, Cambridge CB3 0WA, UK
} 

\author{J. Page}%
%\email{jacob.page@ed.ac.uk}
\affiliation{
 School of Mathematics, University of Edinburgh, EH9 3FD, UK
}

\author{Y. Dubief}
%\email{rrk26@cam.ac.uk}
\affiliation{
 Department of Mechanical Engineering, University of Vermont, Burlington, VT, USA
}

\author{R. R. Kerswell}
%\email{rrk26@cam.ac.uk}
\affiliation{
 DAMTP, Centre for Mathematical Sciences, Wilberforce Road, Cambridge CB3 0WA, UK
}

\date{\today}% It is always \today, today,

\maketitle

\section{Linear stability analysis}

We consider small additive perturbations, $(u^*,v^*,p^*,c_{xx}^*,c_{yy}^*,c_{zz}^*,c_{xy}^*)$
on the steady base state $(U,V,P,C_{xx},C_{yy},C_{zz},C_{xy})$.
%$\phi^*$, to any base state $\Phi$ so that the total field is $\phi=\Phi+\phi^{*}$. 
The base state might correspond to a simple mean flow, e.g. the laminar flow, but also to a more complex 2D state with streamwise variation  such as the saturated PDI travelling waves in the  main text. The linearised equations derived from (1a)-(1c) in the main manuscript are

\begin{widetext}
\begin{subequations}
\begin{align}
    \textit{Re}\ (\partial_t u^{*} + U\partial_x u^* + u^*\partial_x U + V\partial_y u^* + v^*\partial_y U )+\partial_x p^* &= \beta\left(\partial_{xx}u^* + \partial_{yy}u^* \right) + (1-\beta)\left(\partial_x \tau^*_{xx} + \partial_{y}\tau^*_{xy} \right),\\
    \textit{Re}\ (\partial_t v^{*} + U\partial_x v^* + u^*\partial_x V + V\partial_y v^* + v^*\partial_y V )+\partial_y p^* &= \beta\left(\partial_{xx}v^* + \partial_{yy}v^* \right) + (1-\beta)\left(\partial_x \tau^*_{xy} + \partial_{y}\tau^*_{yy} \right),\\
       \partial_t c^*_{xx}+U\partial_x c^*_{xx}+u^*\partial_x C_{xx}+v^{*}\partial_y C_{xx}+V\partial_y c^*_{xx}  +\tau^*_{xx}&=2\left(C_{xx}\partial_x u^*+ c^*_{xx}\partial_x U +  c^*_{xy}\partial_y U  + C_{xy}\partial_y u^{*}\right) \nonumber\\  &\hspace{5cm} +\varepsilon\left(\partial_{xx}c^*_{xx}+\partial_{yy}c^*_{xx}\right),\\
    \partial_t c^*_{yy}+U\partial_x c^*_{yy}+u^*\partial_x C_{yy} + v^{*} \partial_y C_{yy} + V\partial_y c^*_{yy}+\tau^*_{yy}&=2\left(C_{xy}\partial_x v^{*} + c_{xy}^*\partial_x V + C_{yy}\partial_y v^{*} + c_{yy}^*\partial_{y} V\right) \nonumber\\
    &\hspace{5cm} +\varepsilon\left(\partial_{xx}c^*_{yy}+\partial_{yy}c^*_{yy}\right),\\
    \partial_t c^*_{zz}+U\partial_x c^*_{zz}+ u^*\partial_x C_{zz} + v^{*}\partial_y C_{zz} + V\partial_y c^*_{zz}  +\tau^*_{zz}&= \varepsilon\left(\partial_{xx}c^*_{zz}+\partial_{yy}c^*_{zz}\right),\\
    \partial_t c^*_{xy}+U\partial_x c^*_{xy}+ u^*\partial_x C_{xy} + v^{*}\partial_y C_{xy} + V\partial_y c^*_{xy}  +\tau^*_{xy}&=\left(C_{xx}\partial_x v^{*}+ c^*_{xx}\partial_x V + c^*_{yy}\partial_y U+ C_{yy}\partial_y u^{*}\right) \nonumber\\
    &\hspace{5cm} +\varepsilon\left(\partial_{xx}c^*_{xy}+\partial_{yy}c^*_{xy}\right),
    \\
    \partial_x u^{*}+\partial_y v^{*} & = 0,
\end{align}
\label{eq:LS}
\end{subequations}
where 
\begin{subequations}
\begin{align}
    \tau^*_{ij} &= \frac{1}{Wi}\left(\text{tr}(\mathbf{c}^{*})\frac{f^2(\text{tr}\ \mathbf{C})}{L_{\text{max}}^2}C_{ij}+f(\text{tr}\ \mathbf{C}) c^*_{ij}\right),\\
    \partial_k \tau_{ij}^*& =\frac{1}{Wi}\left(\partial_k\text{tr} (\mathbf{c}^*)\frac{f^2(\text{tr}\ \mathbf{C})}{L_{\text{max}}^2}C_{ij} + \text{tr}(\mathbf{c^*})\partial_k \left[\frac{f^2(\text{tr}\ \mathbf{C})}{L_{\text{max}}^2}C_{ij}\right] + 
          c^*_{ij}\partial_k f(\text{tr}\ \mathbf{C}) +  f(\text{tr}\ \mathbf{C})\partial_k c^*_{ij}\right).
\end{align}
\end{subequations}
\end{widetext}
% with $\{i,j\}=\{x,y\}$ and $k=\{x,y\}$. 
%We consider the exponential growth of perturbations and apply a Fourier transform in the homogeneous $x$ direction. 
In the case of laminar flow (steady and independent of $x$) we invoke a normal modes assumption and variation in the streamwise direction is described with a single wavenumber $k$ while the time dependence is exponential with complex growth rate $\sigma := \sigma_p + i \omega_p$. 
% perturbation can be assumed to have exponential time dependence and a single wavenumber $k$ in $x$. 
In this case, the linear stability problem becomes a generalised eigenvalue problem $A\phi^{*}=\sigma B \phi^{*}$ where  $A, B\in \mathbb{C}^{7N_y\times 7N_y}$. If the base flow is spatially periodic (but still steady), Floquet analysis is needed and the linear stability problem has the form of a generalised eigenvalue problem of shape $C\phi^{*}=\sigma D \phi^{*}$, with $C, D\in\mathbb{C}^{7(N_x + 1)N_y\times 7(N_x + 1)N_y}$. 

\begin{figure}
    \centering
    \includegraphics[width=0.9\columnwidth]{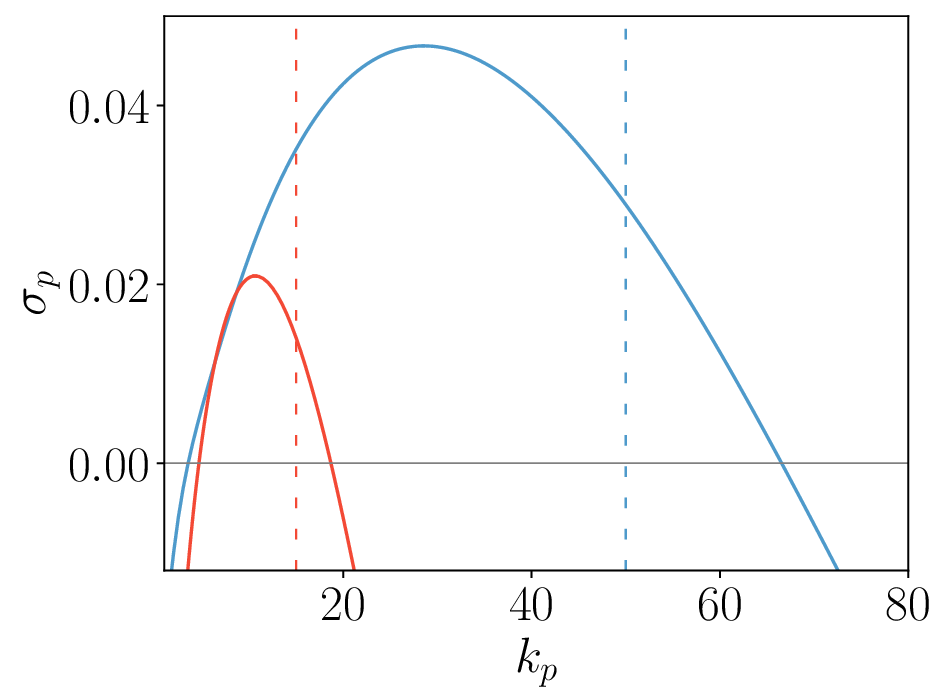}
    \caption{Growth rate of the primary PDI instability, $\sigma_p$, with respect to the fundamental wavenumber $k_p$ for the EIT parameter set $\mathcal{S}_1$ (blue) and the ET parameter set $\mathcal{S}_2$ (red). $\sigma_p>0$ indicates instability. The dashed lines indicate the $k_p$ used in the main text.}
    \label{fig:gr_k}
\end{figure}

\begin{figure}
    \centering
    \includegraphics[width=0.9\columnwidth]{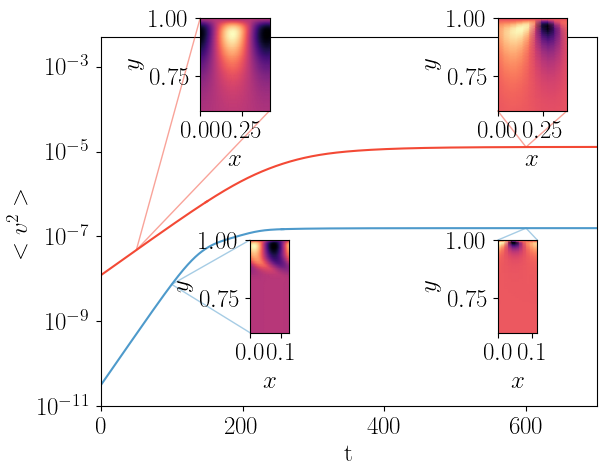}
    \caption{Time evolution of the kinetic energy of the wall-normal velocity for the fundamental PDI with parameter sets $\mathcal{S}_1$ (blue) and $\mathcal{S}_2$ (red). This shows the saturation of the PDI to a finite-amplitude travelling wave. The insets illustrate contours of the wall-normal velocity of the flow at the regions corresponding to the linear instability and the nonlinearly saturated state.}
    \label{fig:timeSeries_PDI}
\end{figure}

\section{The fundamental PDI instability}

We describe here the linear instability (PDI) %corresponding to $k_p^{(1,2)}$ used in the main text and how the base flow for our Floquet analysis is generated.
which saturates to the base flows about which we perform our Floquet analysis.
We consider a computational domain of size $[L_x,L_y]=[\lambda_p^{(1,2)},2]$, where $\lambda_p^{(1,2)}:=2\pi/k_p^{(1,2)}$ and $k_p^{(1,2)}$ is an unstable wavenumber for PDI. 
This choice of domain contains a single wavelength of the primary instability and allows it to develop without becoming unstable itself. Figure \ref{fig:gr_k} shows the growth rate $\sigma_p$ for the primary PDI instability as a function of wavenumber for both parameter sets $\mathcal{S}_{\{1,2\}}$ considered in the main text -- in both cases there is a wide range of unstable wavenumbers. 
In the main text we selected the values $k_p
\in \{50,15\}$ as indicated in figure \ref{fig:gr_k}. 
%since they proved to be capable of sustaining secondary instabilities. 
%The highlighted fundamental instabilities are the launching point to obtain the base flow for our Floquet analysis. 
%
% We chose these particular points after trying a few values of $k_p$ which did not support secondary instabilities.
These particular values of $k_p$ were found to support secondary instabilities (the most unstable waves did not).

We use $[N_x,N_y]=[12,256]$ Fourier-Chebyshev modes in Dedalus \cite{burns2020dedalus} to time step the nonlinear equations (equation 1 in the main text) initialised with the laminar state and low-amplitude $O((10^{-11})$ noise. 
The unstable eigenfunction with wavenumber $k_p$ is excited and grows exponentially according to the linear theory.
This behaviour is confirmed in the early stages of the time series reported in figure \ref{fig:timeSeries_PDI}, which shows the time evolution for the kinetic energy in the wall-normal direction for parameters sets $\mathcal{S}_1$ (blue) and $\mathcal{S}_2$ (red). 
% The exponential growth of the observable for $t\leq 200$ indicates the presence of a linear instability. 
Contours of $v$ for this linear instability mode are also shown in the figure for parameter sets $\mathcal{S}_{1,2}$. Further time evolution of the linear instability results in a nonlinear saturation to a travelling wave, as illustrated by the kinetic energy in the wall-normal direction levelling off (see figure \ref{fig:timeSeries_PDI}). 
% In this case the PDI instability has saturated to a simple travelling wave. 
Contours of $v$ for these simple invariant solution are shown in insets for $t\geq400$ in the same figure. 
The nonlinearly saturated states have similar features to the linear eigenfunctions, but localise further towards the walls.

To construct the base flow used in the Floquet analysis, we converge the saturated states in a Newton-Raphson solver and move to a frame co-moving with the travelling wave where the streamwise velocity is $U = U^{TW} - c$, with $c$ the phase speed of the travelling wave. We then consider linear perturbations to this steady base flow which is solved using the Arnoldi algorithm for sparse matrices (\texttt{scipy.linalg.eigs}) in the \texttt{scipy} Python package \cite{2020SciPy-NMeth} for computational efficiency. The results were checked with an in-house Arnoldi solver. The routines involving sparse eigenvalue problems with ARPACK \cite{lehoucq1998arpack} require the introduction of shift-and-invert modes so that the generalised eigenvalue problem transforms into a simpler one, $C\phi^{*}=\sigma D \phi^{*}$ into $(C-\mu D)^{-1}D\phi^{*} = \sigma^{*}\phi^{*}$ with $\sigma^{*}=1/(\sigma - \mu)$, where $\mu$ is a shift to target the eigenvalues with the largest imaginary parts, i.e. linearly unstable modes.

We timestep our equations in a larger domain to explore the nonlinear evolution of the secondary instabilities. As shown in figure \ref{fig:gr_k} there are several primary instabilities which may grow in the extended domain.  The secondary instability of the flow is explored by perturbing the nonlinearly saturated travelling wave with its eigenfunction resulting from the Floquet analysis. These eigenfunctions might have some projection on the other primary instability modes, but the amplitude of this projection is expected to be much smaller than the projection onto the secondary instability mode. We observed in our simulations that the transition mechanism corresponded to the secondary instability from the Floquet analysis.

\bibliography{apssamp}% Produces the bibliography via BibTeX.